\documentclass[twocolumn,showpacs,aps,prx,amsmath,amssymb,superscriptaddress]{revtex4-2}
\usepackage{amsfonts}
\usepackage{amssymb}
\usepackage{amsmath}
\usepackage[dvips]{graphicx}
\usepackage{color}

\definecolor{nblue}{rgb}{0.3,0.3,1.0}
\definecolor{ngreen}{rgb}{0.2,0.7,0.2}
\definecolor{nred}{rgb}{0.9,0.1,0}
\definecolor{nblack}{rgb}{0,0,0}

\usepackage{amsmath}
\usepackage{graphicx}
\usepackage{amsfonts}
\usepackage{amssymb}
\usepackage[colorlinks=true,linkcolor=blue,urlcolor=blue,citecolor=blue,anchorcolor=blue]{hyperref}
\usepackage{color}
\usepackage{mathrsfs}
\usepackage{isomath}
\usepackage{amsmath}
\usepackage{amsthm}
\usepackage{epstopdf}
\usepackage{txfonts}
\usepackage[textsize=footnotesize]{todonotes}
\usepackage{ulem}

\definecolor{Red}{rgb}{1,0,0}
\definecolor{Black}{rgb}{0,0,0}
\definecolor{Violet}{rgb}{1,0,1}

\definecolor{Orange}{rgb}{1,0.5,0.2}

\definecolor{Bblue}{rgb}{0,0,1}

\newcommand{\Exp}[1]{\langle #1\rangle}

\definecolor{maroon}{rgb}{0.7,0,0}

\begin{document}
		
\title{Certification of non-Gaussian Einstein-Podolsky-Rosen Steering}
		
\author{Mingsheng Tian}
\thanks{M. Tian and Z. Zou contributed equally to this work.}
\address{State Key Laboratory for Mesoscopic Physics, School of Physics, Frontiers Science Center for Nano-optoelectronics, Peking University, Beijing 100871, China}
\author{Zihang Zou}
\thanks{M. Tian and Z. Zou contributed equally to this work.}
\address{State Key Laboratory for Mesoscopic Physics, School of Physics, Frontiers Science Center for Nano-optoelectronics, Peking University, Beijing 100871, China}	
\author{Da Zhang}
\address{Key Laboratory for Physical Electronics and Devices of the Ministry of Education \& Shaanxi Key Lab of Information Photonic Technique, School of Electronic and Information Engineering, Xi'an Jiaotong University, Xi'an 710049, China}
\address{Centre de Nanosciences et de Nanotechnologies C2N, CNRS, Universit\'{e} Paris-Saclay, 91120 Palaiseau, France}
\author{David Barral}
\address{Laboratoire Kastler Brossel, Sorbonne Universit\'{e}, CNRS, ENS-Universit\'e PSL, Coll\`ege de France; 4 place Jussieu, F-75252 Paris, France}
\author{Kamel Bencheikh}
\address{Centre de Nanosciences et de Nanotechnologies C2N, CNRS, Universit\'{e} Paris-Saclay, 91120 Palaiseau, France}
\author{Qiongyi He}
\address{State Key Laboratory for Mesoscopic Physics, School of Physics, Frontiers Science Center for Nano-optoelectronics, Peking University, Beijing 100871, China}
\address{Collaborative Innovation Center of Extreme Optics, Shanxi University, Taiyuan, Shanxi 030006, China}
\address{Peking University Yangtze Delta Institute of Optoelectronics, Nantong, Jiangsu, China}	
\address{Hefei National Laboratory, Hefei 230088, China}	
\author{Feng-Xiao Sun}
\email{sunfengxiao@pku.edu.cn}
\address{State Key Laboratory for Mesoscopic Physics, School of Physics, Frontiers Science Center for Nano-optoelectronics, Peking University, Beijing 100871, China}
\address{Collaborative Innovation Center of Extreme Optics, Shanxi University, Taiyuan, Shanxi 030006, China}
\author{Yu~Xiang}
\email{xiangy.phy@pku.edu.cn}
\address{State Key Laboratory for Mesoscopic Physics, School of Physics, Frontiers Science Center for Nano-optoelectronics, Peking University, Beijing 100871, China}
\address{Collaborative Innovation Center of Extreme Optics, Shanxi University, Taiyuan, Shanxi 030006, China}
		
\begin{abstract}		
Non-Gaussian quantum states are a known necessary resource for reaching a quantum advantage and for violating Bell inequalities in continuous variable systems. As one kind of manifestation of quantum correlations, Einstein-Podolsky-Rosen (EPR) steering enables verification of shared entanglement even when one of the subsystems is not characterized. However, how to detect and classify such an effect for non-Gaussian states is far from being well understood.
Here, we present an efficient non-Gaussian steering criterion based on the high-order observables and conduct a systematic investigation into the hierarchy of non-Gaussian steering criteria. 
Moreover, we apply our criterion to three experimentally-relevant non-Gaussian states under realistic conditions and, in particular,  propose a feasible scheme to create multi-component cat states with tunable size by performing a suitable high-order quadrature measurement on the steering party. 
Our work reveals the fundamental characteristics of non-Gaussianity and quantum correlations, and offers new insights to explore their applications in quantum information processing.
\end{abstract}
		
\maketitle

\section{Introduction}

In 1935, Einstein, Podolsky and Rosen (EPR) pointed out one of the most counter-intuitive features of quantum mechanics, that measurements on one particle $A$ seem to immediately affect the state of the other distant particle $B$~\cite{eprparadox1935}. Schrödinger then introduced the term ‘steering’ for such nonlocal effect manifested by the EPR argument~\cite{schrodinger1935}. Since it was reformulated with an operational definition in 2007~\cite{wiseman2007}, where two parties can verify shared entanglement even if one measurement device is untrusted~\cite{rmp-2009,review-steering-2016,review-steering-rmp2020,xy-prxquantum2022}, EPR steering is recognized as an essential resource for a number of one-sided device-independent quantum information processing tasks, such as quantum key distribution~\cite{QKD2014np,QKD2015nc,QKD-OPTICA}, quantum secrete sharing~\cite{he2015np,secretsharingpra2017}, and randomness generation~\cite{randomness-njp,randomness-prl2018,randomness-prl2019}. 

In the quest to demonstrate a quantum advantage in the continuous-variable (CV)~\cite{PhysRevLett.109.230503,PhysRevResearch.4.043010}, recently, substantial progress for controllable generation of non-Gaussian entangled states has been made in various platforms, e.g., photon-subtracted Gaussian states~\cite{np11.713,lsh-prl}, spontaneous parametric down-conversion systems~\cite{pra2015,spdc2004,spdcprx2020}, and microwave cavities with superconducting artificial atoms~\cite{wang2016schrodinger,wang2022flying,PRXQuantum.2.030204}. Such states usually carry higher-order statistical moments of the quadrature field operators beyond Gaussian and therefore have been proven to be indispensable resource for CV entanglement distillation~\cite{eisert2002distilling,fiuravsek2002gaussian,nphoton.2010.1,nha-scirep2016}, quantum-enhanced sensing~\cite{science.345,PhysRevLett.107.083601} and imaging~\cite{PhysRevApplied.16.064037,mattia}. In particular, it was proven that a necessary requirement for the sampling complexity of bosonic quantum computation is the existence of non-Gaussian entanglement~\cite{PhysRevLett.130.090602}. However, the characterization of entanglement in general non-Gaussian systems, 
especially when some of the parties
are considered untrusted, still has a number of conceptual and practical challenges.

The major difficulty in certifying non-Gaussian entanglement is the nontrivial correlations appearing in higher-order moments of the observables which cannot be sufficiently uncovered by most of the existing criteria based on the second-order correlations~\cite{simon,van2003,gerardo2005,hz2006,reid2014pra,he2015prl}. 
To tackle this problem, a few approaches have been developed to witness quantum inseparability for non-Gaussian states, such as the generalized Hillery-Zubairy (HZ) criteria~\cite{hzspdc2020}, the nonlinear entanglement criteria based on amplitude-squared squeezing~\cite{reid2015,zhangda2021prl}, and criteria based on quantum Fisher information or metrological squeezing parameters~\cite{PRXQuantum.3.030347,NC2021,tian2022praaplied}.
As steerability is a more
stringent form of correlations than inseparability, thus, the question of how to detect it in non-Gaussian schemes naturally arises and is crucial to explore its further applications in quantum computation protocols.

In this work, we propose an efficient  non-Gaussian steering criterion based on the covariance matrix (CM) derived from experimentally accessible high-order observables, and systematically compare it with the other two sufficient criteria which are referred as HZ- and Reid-type criteria. 
We show that 
both the CM-type and linear Reid (LR)-type criteria, which utilize linear estimation, are equally effective in detecting EPR steering for general non-Gaussian states, and are better than HZ-type criterion.
The high-order quadratures in non-Gaussian states usually stimulate more complicated nonlinear correlations, which may not be captured by LR-, CM- and HZ-type criteria. Therefore, we further use a nonlinear estimation to acquire the optimal correlations after making an arbitrary-order measurement on the steering party. 
We apply our findings to a series of experimentally-relevant non-Gaussian states, including spontaneously down-converted triple photons ~\cite{spdc2004,spdcprx2020}, photon-subtracted Gaussian states~\cite{subtraction2007science,nicolasnp2020,lsh-prl}, and non-Gaussian mixtures of Gaussian states. We observed that all three criteria successfully detect non-Gaussian steering with high-order quadratures but their performance in terms of the system parameters is different, which aligns with our results (\textit{i.e.}, CM=LR$\geq$ HZ).
Additionally, we demonstrate that the high-order quadratures used for our criterion can also be utilized to produce multi-component cat states with tunable size, which are very promising candidates for fault-tolerant computation~\cite{error-2014,error-PhysRevLett.119.030502,errorcorrection2021}.

\section{High-order quadratures for non-Gaussian systems}

Suppose that two spacelike separated observers, say Alice and Bob, share a bipartite physical state described by annihilation operators $\hat{a}$ and $\hat{b}$, respectively. When the shared two-mode state with density operator $\hat{\rho}$ is a Gaussian state, all quantum-correlation properties of the state are synthesized in a single $4 \times 4$ real symmetric matrix --the covariance matrix $V$ -- with elements $V_{ij}=\frac{1}{2}\Exp{\hat{\xi}_i\hat{\xi}_j+\hat{\xi}_j\hat{\xi}_i}-\Exp{\hat{\xi}_i}\Exp{\hat{\xi}_j}$ where $\hat{\xi}=(\hat{X}_{A},\hat{P}_{A},\hat{X}_{B},\hat{P}_{B})^{\top}$. $\hat{X}_O=(\hat{o}+\hat{o}^{\dagger})/2$,
$\hat{P}_O=-i (\hat{o}-\hat{o}^{\dagger})/2$ are respectively amplitude and phase quadrature operators of the mode $O=A,B$ ($\hat{o}=\hat{a},\hat{b}$) ~\cite{simon,boundentanglement}.

But if Alice and Bob share a non-Gaussian state, as more complex correlations may exist -- beyond second-order moments, the CM fails in general to fully describe the quantum correlations of the system. High-order (HO) quadratures are thus necessary to depict the features of high-order correlations. In order to characterize these HO quantum correlations, we introduce the $k$-th order amplitude and phase quadratures $\hat{X}_O^{k}=(\hat{o}^{k}+\hat{o}^{\dagger k})/2$ and $\hat{P}_O^{k}=-i (\hat{o}^{k}-\hat{o}^{\dagger k})/2$. For instance, for $k=1$ we recover the standard quadrature operators, and for $k=2$ the new quadratures are the real and imaginary part of squared-amplitude operators and their eigenstates are Schrödinger cats~\cite{hillerypra1987}. 

The covariance matrix $V^{kl }$ of a bipartite system which synthesizes the quantum correlations related to these HO quadratures is defined as
$V^{kl}_{ij}=\frac{1}{2}\Exp{\hat{\xi}^{kl}_i\hat{\xi}^{kl}_j+\hat{\xi}^{kl}_j\hat{\xi}^{kl}_i}-\Exp{\hat{\xi}^{kl}_i}\Exp{\hat{\xi}^{kl}_j}$, where $\hat{\xi}^{kl}=(\hat{X}_{A}^k,\hat{P}_{A}^k,\hat{X}_{B}^l,\hat{P}_{B}^l)^{\top}$ with $k,l=1, 2, \dots$. With this CM, we can analyze for instance the quantum correlations between standard quadratures in Alice's system and square-amplitude quadratures in Bob's ($k=1, l=2$). For $k=l=1$ we recover the Gaussian CM and the correlations between standard quadratures. Our HO covariance matrix fulfills the following compact form of the uncertainty principle for arbitrary chosen orders $k$ and $l$ ~\cite{njp-entanglement,cm-high,Nha}:
\begin{equation}\label{ur}
   V^{kl}+\frac{i}{2}( J_{A}^k\oplus J_{B}^l)\geq 0,
\end{equation}
where $J_{O}^{k}=\text{adiag}(i Q_{O},-i Q_{O})$ is a generalized $2\times2$ symplectic matrix for the mode $O$ and $Q_{O}=\langle[\hat{o}^{k}, \hat{o}^{\dag k}]\rangle/2$.

\section{Criteria for Non-Gaussian Steering}\label{sec:criteria}
From a quantum information perspective, EPR steering can be formalized in terms of entanglement verification when one of the parties is untrusted, or has uncharacterized devices~\cite{wiseman2007}. Mathematically, Alice steers Bob if and
only if the probabilities of all possible joint measurements implemented on the bipartite state $\hat{\rho}$ cannot be factorized into a local hidden state model of the form
\begin{equation}\label{lhs}
\mathcal{P}(\alpha,\beta|\hat{A},\hat{B},\hat{\rho})=\sum_{\zeta}\mathcal{P}_{\zeta}\mathcal{P}(\alpha|\hat{A},\zeta)\mathcal{P}(\beta|\hat{B},\hat{\rho}_{\zeta}),~~~\forall \hat{A},\hat{B},
\end{equation}
where the hidden parameter is described by a random variable $\zeta$, $\{\hat{\rho}_{\zeta}\}$ is an ensemble of marginal states for Bob, and $\alpha,\beta$ denote the outcomes of local observables $\hat{A}$ and $\hat{B}$ for Alice and Bob, respectively. However, the above definition of EPR steering is not experimentally efficient in general as one needs full knowledge of the state, and in particular for CV non-Gaussian states whose density matrix is usually very hard to obtain.

To overcome this relevant problem, a series of criteria can be introduced to provide a more direct and experimentally-friendly characterization of non-Gaussian steering based on HO quadratures in phase space. Below, we present three different criteria for non-Gaussian steering grounded on HO quantum correlations: the covariance matrix type, the Hillery-Zubairy type and the Reid type; and we introduce the steering parameter $S$ as a common definition of steering in order to compare them.

\subsection{CM-type criterion}
First, we propose a EPR steering criterion based on the CM composed of correlations between high-order quadratures.
For a physical quantum state, it has been proved that the high-order covariance matrix $V^{kl}$ still fulfill the uncertainty principle (\ref{ur}) for chosen orders $k$ and $l$~\cite{simon,zhangda2021prl,Nha}. Hence the state is $A\rightarrow B$ steerable if the condition 
\begin{equation}\label{eqcm}
  M_{B|A}^{kl}=V^{kl}+\frac{i}{2}( 0_{A}\oplus J_{B}^l)\geq 0
\end{equation}
is violated. The proof is as follows. Suppose the state is not $A\rightarrow B$ steerable, then it must be characterized by a local hidden state model in the form of Eq.~(\ref{lhs}), which leads to the local uncertainty of Alice's subsystem $J_A^k$ becomes null~\cite{eprcm-pra2014}. Define the $2\times 2$ real symmetric covariance matrices $V_A^k$ and $V_B^l$ as the CMs corresponding to the reduced states of Alice’s and Bob’s subsystems, satisfying $V_A^{k}\geq 0$ and $V_B^{l}+\frac{i}{2}J_B^{l}\geq 0$, respectively. Therefore, Eq.(\ref{eqcm}) can be derived.

Here, we denote the minimal eigenvalue of the left hand of the above inequality~(\ref{eqcm}) as $S_{\rm{CM}}$. 
Now subsystem $B$ is steerable by subsystem $A$ when
\begin{equation}\label{eqduan}
S_{\rm{CM}} < 0.
\end{equation}

\subsection{HZ-type Criterion}
The Hillery and Zubairy criterion was originally proposed to detect entanglement~\cite{hz2006}, and then generalized to EPR steering for the linear quadrature operators $\hat{X}$ and $\hat{P}$ ($k=l=1$)~\cite{hzhe2011,hz-he2012}. 
For non-Gaussian systems, the high-order operators ($k, l >1$) can also be taken into account~\cite{hz-he2012}, hence the HZ-type criterion reads
\begin{equation}\label{HZ}
H_{B|A}^{kl}=|\langle \hat{A}\hat{B} \rangle |-\sqrt{ \langle \frac{\hat{A}^{\dagger }\hat{A}+\hat{A}\hat{A}^{\dagger } }{2}\rangle
\langle \hat{{B}^{\dagger }}\hat{B} \rangle}\leq 0,
\end{equation}
where we introduced the operators $\hat{A}=\hat{a}^k-\langle \hat{a}^k \rangle$ and $\hat{B}=\hat{b}^l-\langle \hat{b}^l \rangle$.
When inequality~(\ref{HZ}) is violated, the system is said to be "steerable".
To unify the bound of different criteria, here we redefine the steering parameter as $S_{\rm{HZ}}:=-H_{B|A}^{kl}$. Thus, the subsystem $B$ is steerable by subsystem $A$ if
\begin{equation}\label{eprhz}
S_{\rm{HZ}} < 0.
\end{equation}

\subsection{Reid-type Criterion}
In 1989, Reid introduced a criterion that identifies an EPR paradox corresponding to a violation of the Heisenberg uncertainty relation \cite{reid1989demonstration}. 
This kind of criterion can also be generalized for non-Gaussian steering by involving high-order quadratures~\cite{reid2015}, which takes the form of
\begin{equation}
R_{B|A}^{kl}=\frac{2\sqrt{\langle V_{A}^{inf}\left(\hat{X}_B^l \right)\rangle \langle V_{A}^{inf}\left(\hat{P}_B^l \right)\rangle}}{|\langle[\hat{X}_B^l,\hat{P}_B^l]\rangle|}<1.
\end{equation}
Here, $\langle V_{A}^{inf}  \left( \hat{O}_B^l \right) \rangle =\sum_a P_{a|\hat{O}_A^k}  V_{a|\hat{O}_A^k}^{inf}\left(\hat{O}_B^l \right)$ is the inference variance, and $V_{a|\hat{O}_A^k}^{inf}\left(\hat{O}_B^l \right)$ is calculated from the conditional state ${\hat{\rho}}_{a|\hat{O}_A^k}$ by measuring a high-order quadrature $\hat{O}_A^k$ where $\hat{O}=\hat{X}$ or $\hat{P}$. For sake of comparison, here we redefine the steering parameter as $S_{\rm{CR}}:=R_{B|A}^{kl}-1$. 
Now subsystem $B$ is steerable by subsystem $A$ if
\begin{equation}\label{con}
S_{\rm{CR}}< 0.
\end{equation}

To avoid the measurement of conditional distributions, a practical linear estimation is usually applied to derive the inference variance 
$\langle V_{A,LR}^{inf}\left( \hat{O}_B \right) \rangle=\mathrm{\Delta}^2(\hat{O}_B+g_o \hat{O}_A)$ 
with a gain factor $g_o=-\langle \hat{O}_B,\hat{O}_A\rangle/\mathrm{\Delta}^2(\hat{O}_A)$ and $\langle \hat{O}_B,\hat{O}_A\rangle=(\langle \hat{O}_B\hat{O}_A\rangle+\langle \hat{O}_A\hat{O}_B\rangle)/2-\langle \hat{O}_B\rangle\langle\hat{O}_A\rangle$~\cite{reid1989demonstration}. 
For a Gaussian state, this kind of linear estimation when $k=l=1$ is the optimal strategy to minimize the quantity  $\langle V_{A}^{inf}  \left( \hat{O}_B^l \right) \rangle$~\cite{reid1989demonstration,rmp-2009}, which means the criterion with linear estimation can necessarily and sufficiently detect Gaussian steering. 
But when the state is non-Gaussian, the linear estimation is usually not the optimal one, which can be corroborated by the examples considered in the next section. Similarly, we label the Reid-type criterion with linear estimation as $S_{\rm{LR}}:=R_{B|A,LR}-1$. The steering from mode $A$ to mode $B$ can be characterized by
\begin{equation}
S_{\rm{LR}}<0. 
\end{equation}

\section{Hierarchy of non-Gaussian steering criteria}\label{sec:hierarchy}
The non-Gaussian steering criteria introduced in the previous section are rooted in different grounds and present singular features. Thus, one can wonder which criterion is the strongest. Previous works investigated the relations among different criteria~\cite{equiv-tan,equiv-zss,eqiv-spin}. However, those are based on second-order correlations, and hence only suitable in all-Gaussian scenarios.

Here, we provide a rigorous proof in the form of two theorems and a corollary that shows a hierarchy in these four non-Gaussian steering criteria and demonstrates that their abilities to detect non-Gaussian steering follows the relation 
\begin{equation}\nonumber
\rm{CR} \geq \rm{LR}=\rm{CM}\geq \rm{HZ}.
\end{equation}

\textbf{Theorem 1.}$-$ 
The LR-type steering criterion is equivalent to the CM-type criterion, and both of them can detect steering in a larger range than the HZ-type criterion.

\textit{Proof.}$-$
For each pair of orders $k$ and $l$ related respectively to Alice and Bob's modes, the CM of any physical state can be rewritten in standard form as~\cite{simon}
\begin{equation}\label{cv}
V^{kl}=\left(\begin{array}{cccc}
n & 0 & c_{1} & 0 \\
0 & n & 0 & c_{2} \\
c_{1} & 0 & m & 0 \\
0 & c_{2} & 0 & m
\end{array}\right).
\end{equation}

The LR-steering criterion can be expressed in terms of the elements of the standard-form CM using suitably-optimized gains $g$. The steerability from Alice to Bob is then equivalent to (see appendix \ref{appA})
\begin{equation}\label{wlr}
    W_{\rm{LR}}\equiv [n m - c_{1}^{2}][ n m - c_{2}^{2}] -(n Q_{B})^2 <0,
\end{equation} 
where $Q_{B}=\langle[\hat{B},\hat{B}^\dagger]\rangle/4$.

To prove the equivalence of the CM-type and the LR-type criteria we first demonstrate that if the LR-type criterion witnesses non-Gaussian steering the CM-type also witnesses it. From Eqs.~(\ref{eqcm}) and (\ref{wlr}), we find the relation $W_{\rm{LR}}=\mathrm{det}(M^{kl}_{B|A})$ (see appendix \ref{appA}). Thus, $W_{\rm{LR}}<0$ implies that the matrix $M^{kl}_{B|A}$ is not positive semi-definite, which matches the steering condition for the CM-type criterion.

Now we need to demonstrate the reverse case, \textit{i.e.}, if the CM-type criterion can witness steering, the LR-type can also witness it in the same range of parameters.
That the CM-type criterion witnesses steering means that the $M^{kl}_{B|A}$ is not positive semi-definite, which indicates that there must be at least one negative determinant of all upper-left $s\times s$ ($s=1, 2, 3, 4$) corners of $M^{kl}_{B|A}$, where the determinant is called as the leading principal minor of order $s$~\cite{linearalgebra}. 
From the definitions of $M^{kl}_{B|A}$ and  $V^{kl}$, Eqs.~(\ref{eqcm}) and (\ref{cv}), we find that their leading principal minor of order $s\leq3$ are the same. As the covariance matrix $V^{kl}$ for any physical state must be positive semi-definite, all the upper-left $s\times s$ corners of $V^{kl}$ must have non-negative determinants. Thus, a non-positive semi-definite $M^{kl}_{B|A}$ only exists when its $4$th-order leading principal minor is negative, \textit{i.e.}, $\mathrm{det}(M^{kl}_{B|A})<0$, which is exactly the condition of the LR-type criterion. Thus both LR- and CM-type criteria are equivalent.

Next, we prove that the LR-type criterion is stronger than the HZ-type criterion. The HZ-type criterion of Eq. \eqref{HZ} in terms of the elements of the standard-form CM given by Eq. \eqref{cv} states that mode $A$ can steer mode $B$ if
\begin{equation}\nonumber
    W_{\rm{HZ}}\equiv [n m -\frac{(c_1-c_2)^2}{4}]^2 -(n Q_{B})^2<0.
\end{equation}
In order to find the relation between LR- and HZ-type criteria, we notice that 
\begin{equation}\label{lrhz}
        W_{\rm{LR}}-W_{\rm{HZ}}=-\frac{(c_1+c_2)^2}{16}\left[8 n m - 4 c_1 c_2 +(c_1-c_2)^2\right].
\end{equation}
Since all principal minors of the CM $V^{kl}$ are non-negative one finds that $n m \geq (c_{1(2)})^2$. Thus, the part between brackets in Eq. \eqref{lrhz} is strictly positive. This leads to the following result
\begin{equation}
    W_{\rm{LR}}-W_{\rm{HZ}}\leq 0.
\end{equation}
This expression indicates that if the HZ-type criterion detects non-Gaussian steering, the LR-type criterion would detect it as well. In other words, the LR-type criterion is stronger than the HZ-type criterion. \qed


\textbf{Corollary}.$-$ The CR-type steering criterion, which is based on conditional states instead of linear estimation, has a larger range than the others.

\textit{Proof.}$-$ This is a direct consequence of fact that the LR-type criterion is a subcase of the CR-type criterion such that $\rm{CR}\geq \rm{LR}$. Then, by Theorem 1 we have: $\rm{CR}\geq \rm{LR}=\rm{CM}\geq \rm{LZ}$. \qed


Now we check the relation among the four criteria for an experimentally-relevant case: the condition $c_1=-c_2$. This relation is equivalent to $\langle \hat{A}\hat{B}^{\dagger}\rangle=\langle \hat{a}^k(\hat{b}^{\dagger})^l\rangle-\langle \hat{a}^k\rangle\langle (\hat{b}^{\dagger})^l\rangle=0$ (see appendix \ref{appA}). Such condition is fulfilled by widely used two-mode CV states including the Gaussian EPR state with phase-insensitive losses and the two-mode squeezed thermal state ($k=l=1$)~\cite{eprstate}, and the spontaneously down-converted triple photon state ($k=1$ and $l=2$)~\cite{spdc2004,spdcprx2020}. These states are preferred resources for various CV protocols, like quantum teleportation and error correction in CV quantum computing. We have the following theorem for this case:

\textbf{Theorem 2.}$-$ The CM-, HZ-, and LR-type steering criteria are equivalent if the correlation elements $c_{1,2}$ of the standard-form CM of Eq. (\ref{cv}) satisfy the relation $c_1=-c_2$.

\textit{Proof.}$-$ We have proved above that the LR- and CM-type criteria are equivalent. 
Here we only need to prove the equivalence between the LR-type and HZ-type criteria. 
With $c_1=-c_2$, $W_{\rm{LR}}-W_{\rm{HZ}}\equiv 0$ in Eq.~(\ref{lrhz}), which means that the HZ-type criterion is equivalent to the LR-type criterion. This completes the proof. \qed  

\begin{figure}[b]
    \centering
    \includegraphics[width=8.5cm]{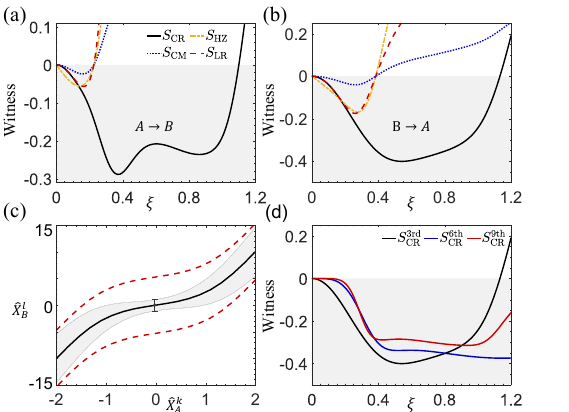}
    	\caption{(a)(b) Evolution of different types of EPR steering criteria introduced in Sec.~\ref{sec:criteria} as function of the effective coupling strength $\xi=\alpha_p g t$ in partially-degenerate three-photon SPDC. A negative value of the witness reveals the existence of steering from subsystem $A$ to subsystem $B$ (a) or from subsystem $B$ to subsystem $A$ (b).
    	(c) Quadratic quadrature conditional joint probability distributions between the subsystems $A$ and $B$. Mean value of the quadratic quadrature of subsystem $B$ conditioned on the linear quadratures of subsystem $A$ (black trace). The gray shaded region displays the normalized standard deviations of the distribution. 
     The red dashed curves give the corresponding normalized standard deviations by linear estimation of $\hat{X}_A^k$ and $\hat{X}_B^l$ ($k=1$ and $l=2$) . 
     Here we fix $\xi=1.2$. 
    (d) Evolution of the different-order of CR-type steering witness. The superscript ``3rd" indicates $k=1$, $l=2$, ``6th" is $k=2$, $l=4$, and ``9th" is $k=3$, $l=6$. In all figures we take $\alpha_p=5$.}
    	\label{fig1}
\end{figure}

\section{Detecting Non-Gaussian steering in Practical Systems}
Very recently, non-Gaussian quantum states have been generated  through suitable nonlinear interactions~\cite{spdc2004,science.345,spdcprx2020}, or applying  appropriate non-Gaussian operations on  Gaussian states, such as single-photon addition/subtraction ~\cite{subtraction2007science,nicolasnp2020}, etc. In the following, we exhibit the behaviors of the above criteria in three typical non-Gaussian systems, \textit{i.e.}, three-photon spontaneous parametric down-conversion (SPDC) states, photon-subtracted Gaussian states, and non-Gaussian mixtures of Gaussian states.

\subsection{Three-Photon SPDC system}

We begin with a typical instance of the third-order nonlinear interaction process, where the non-Gaussian state is generated by a three-photon SPDC process with two modes degenerated. 
In such a system, a pump photon at frequency $\omega_P$ is down converted to a signal photon at frequency $\omega_A$ and two degenerate idler photons at frequency $\omega_B$, with the energy conservation relation $\omega_P=\omega_A+2\omega_B$. This process is described by the Hamiltonian
\begin{equation}\label{eqhab2}
	\hat{H}=i\hbar g \hat{a}^{\dagger}\hat{b}^{\dagger 2}\hat{p}+H.c., 
\end{equation}
where $g$ is the third-order coupling strength of the nonlinear interaction, $\hat{a}$ and $\hat{b}$ are respectively the annihilation operators of the down-converted signal mode $A$ and idler mode $B$, and $\hat{p}$ is the annihilation operators of the pump. 
Here, we consider the initial state $\hat{\rho}_0$ as a vacuum state with a fixed coherent pump $|\alpha_p\rangle$.
 
Due to the specific form of the Hamiltonian~(\ref{eqhab2}), the covariance matrix with the first-order quadratures cannot capture the correlations between two modes~\cite{reid2015}.
To witness the non-Gaussian steering in such state, we now refer to the criteria mentioned above and introduce high-order quadratures $\hat{X}_A^k$, $\hat{X}_B^l$ with $l=2k$.
Thus, the CM obtained by $V^{kl}_{ij}=\langle\{\Delta\hat{\xi}^{kl}_{i},\Delta\hat{\xi}^{kl}_{j}\}\rangle_{\hat{\rho}}$ with $\hat{\rho}=e^{-i\hat{H}t}\hat{\rho}_0 e^{i\hat{H}t}$ possesses a standard form with $c_1=-c_2$ in Eq.~(\ref{cv}).
Figs.~\ref{fig1}(a) and (b) show all three types of the EPR steering witness constructed in Sec.~\ref{sec:criteria} as a function of the effective coupling strength $\xi=\alpha_p g t$. 
For $k=1$ and $l=2$, the four criteria witness non-Gaussian steering in a given coupling strength range. Particularly, CM- (blue dots), HZ- (orange dashdotted), and LR-type (red dashed) criteria intersect with the steering witness bound at the same point, demonstrating that the three witnesses have the same ability to detect non-Gaussian steering as established in Theorem 2.

A main result of section IV is that the CR-type steering criterion has a larger detection range than the other three. Here, the CR-type steering criterion captures the nonlinear relation between the squared-amplitude quadrature $\hat{X}_B^2$ of subsystem $B$ and the linear quadrature $\hat{X}_A^1$ of subsystem $A$, as shown in Fig.~\ref{fig1}(c) (black trace). It follows then that the LR(CM)-type criterion, which only contains linear relations between HO quadratures, cannot capture such nonlinear correlations. Thus, the CR-type criterion, which is rooted in the conditional state after performing HO quadrature measurements, takes an obvious advantage in comparison with others.
To demonstrate this point, we computed the variance of the optimized linear combinations in LR-type, expressed as $\mathrm{\Delta}^2(\hat{X}_B^k+g_x \hat{X}_A^l)$. In Fig.~\ref{fig1}(c), the variance resulting from linear estimation (red dashed curves) is consistently greater than the corresponding variance resulting from nonlinear estimation (shaded region). This observation suggests that nonlinear estimation can capture more correlations via lower variance, thereby enabling the CM-type criterion to outperform the other three.

\begin{figure}[tb]
    \centering
    \includegraphics[width=8.5cm]{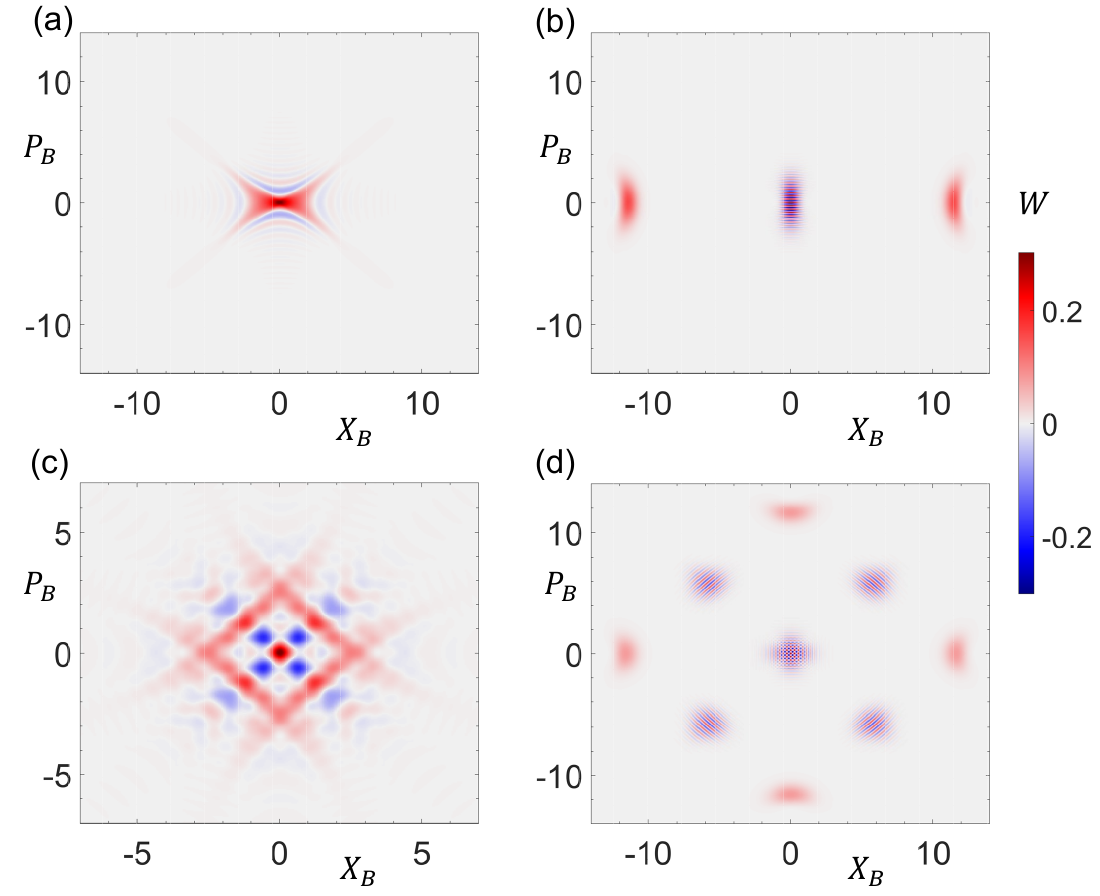}
    	\caption{Wigner functions of the conditional states on Bob's side when Alice performs a  quadrature measurement $\hat{X}_A^k$ with different orders. For $k=1$, a kitten state (a) and a large cat state (b) are obtained on Bob' side for values of measured quadratures of $X_A^1=0.6$ and $X_A^1=6$ on Alice' side, respectively. In comparison, when Alice performs a high-order quadrature measurement $\hat{X}_A^2$ on her side, a four-component kitten state (c) and a large cat state (d) are generated for measured values of $X_A^2=1.3$ and $X_A^2=34.5$, respectively. Here, we take the interaction strength $\xi=0.6$ and the pumping amplitude $\alpha_p=5$.}
    	\label{fig2}
\end{figure}

Besides, we find that the nonlinear steering disappears when increasing the coupling strength $\xi$ even for the CR-type criterion [solid black lines in Figs.\ref{fig1}(a) (b)].
However, from the Hamiltonian (\ref{eqhab2}) the correlations should be enhanced with the interaction strength~\cite{reid2015}. 
The reason is that quadratures $k=1$ for mode $A$ and $l=2$ for mode $B$, considered in Figs.~\ref{fig1}(a) and (b), only reflect part of correlated information.
In order to capture the full information on the correlations, we construct higher-order quadratures with $k=2$, $l=4$, and $k=3$, $l=6$, and plot the evolution of the corresponding CR-type steering witness in Fig.~\ref{fig1}(d).
With the increasing of $\xi$, the steering carried by ${S}_{\rm{CR}}^{3rd}$ (corresponds to $k=1$ and $l=2$) disappears, whereas ${S}_{\rm{CR}}^{6th}$ ($k=2$, $l=4$) and ${S}_{\rm{CR}}^{9th}$ ($k=3$, $l=6$) still survive. This feature is significantly different from Gaussian states generated by two-photon SPDC, in which Gaussian entanglement captured by first- and seconded-order moments always exists with an increasing effective coupling strength (\textit{i.e.}, squeezing parameter) \cite{gerardo2005}.
In contrast, for the non-Gaussian states generated by three-photon SPDC, higher-order quadratures are required to witness the entanglement information.

The CR-type criterion relies on the high-order quadrature measurements performed on the untrusted side. For non-Gaussian states, this operation can not only be used to witness EPR steering, but also to generate Schr\"{o}dinger cat states~\cite{reid2015,sunfengxiao2021prl}.
Comparing with other approaches ~\cite{cat2006,cat2007,cat2013}, high-order quadrature measurements allow to tune continuously the size of the cat state due to its continuous-variable nature. We plot in Figs.~\ref{fig2}(a) and (b) the Wigner  functions of the projected modes $\hat{\rho}_B$ considering a quadrature measurement ($k=1$) on Alice's side.
For a small value $X_A^{1}=0.6$ we obtain a two-component kitten state $|\alpha \rangle+|-\alpha \rangle$ of amplitude $\alpha=1$, whereas for a large value $X_A^{1}=6$, the same state with amplitude $\alpha=8$ is observed.

Interestingly, if the quadrature measurement performed by Alice is high-order with $k=2$, a four-component cat state $|\alpha\rangle +|-\alpha\rangle - |i \alpha\rangle - |-i \alpha\rangle$  is obtained ~\cite{error-2014}. This is shown in Figs.~\ref{fig2}(c) and (d). Again, the size of the four-component cat state can be tuned continuously with different measurement outcomes on Alice's side. The fidelity between the generated state for $X_A^2=1.3$ and an ideal cat state with $\alpha=1.8$ is $0.922$ (Fig.~\ref{fig2}(c)), and the fidelity between the generated state for $X_A^2=34.5$ and an ideal cat state with $\alpha=8$ is $0.938$ (Fig.~\ref{fig2}(d)). Remarkably, these four-component cat states are a resource for fault-tolerant quantum computation~\cite{error-2014,error-PhysRevLett.119.030502,errorcorrection2021}. Thus, our non-Gaussian steering scheme can be a key element in future quantum information processing in fields like optics, optomechanics or superconducting systems.

\subsection{Photon-subtracted Gaussian states}

\begin{figure}[tb]
    \centering
    \includegraphics[width=8.5cm]{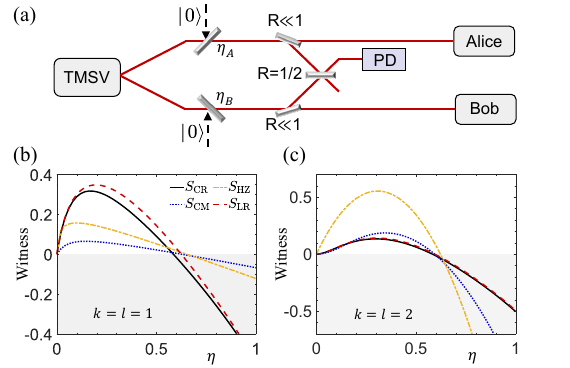}
    \caption{(a) An experimental scheme to generate non-Gaussian states by performing coherent single-photon subtraction on a lossy TMSV state. The generated non-Gaussian state is selected and conditioned on registering a photon at the photon-detector (PD). Evolution of different types of steering witness are given in (b) for $k=l=1$, and in (c) for $k=l=2$. Here, $\eta_A=\eta_B=\eta$ is the transmission efficiency of Alice's and Bob's channels. The squeezing parameter of the initial TMSV state is fixed at $r=1$, and the gray shades represent the regions of detected non-Gaussian EPR steering.
	}
    \label{fig3}
\end{figure}
Another common approach to generate non-Gaussian states is performing a single-photon subtraction to an initial Gaussian state~\cite{subtraction-np2009,subtraction-PhysRevLett.98.030502}. 
It can be realized with the protocol presented in Fig.~\ref{fig3}(a), where an optical parametric amplifier (OPA) produces a two-mode squeezed vacuum state (TMSV),
$
	|\psi_{\mathrm{TMSV}}\rangle_{AB}\propto\sum^{\infty}_{m=0}\tanh^{m}r |m,m\rangle_{AB}.
$
Here, $r$ is the squeezing parameter and $|m\rangle$ is the Fock basis. 
We apply a coherent single-photon subtraction operation on the two modes. If the photon-detector successfully detects a photon, the full density matrix becomes  $\hat{\rho}_{\mathrm{loss}}^{coh} =(\hat{a}+\hat{b}) \hat{\rho}_{\mathrm{loss}}(\hat{a}^{\dagger}+\hat{b}^{\dagger})$, where $\hat{\rho}_{\mathrm{loss}}$ is the initial joint density matrix $|\psi_{\mathrm{TMSV}}\rangle \langle \psi_{\mathrm{TMSV}}|_{AB}$ after transmission losses~\cite{xiangyu2017}. These losses are characterized by the transmission efficiency of each channel $\eta_{A,B}$.

We plot in Fig.~\ref{fig3} (b) all the mentioned steering criteria for this state with respect to the channel efficiency $\eta_A=\eta_B=\eta$ using first-order quadratures ($k=l=1$), where the non-Gaussian steering is always detected if there is no channel loss ($\eta=1$). Here, different criteria exhibit different resilience to channel loss: the CR-type is the most robust one, and only requires a channel efficiency of $\eta>0.585$, while both the LR- and CM-type requires $\eta>0.62$. The HZ-type, however, performs worst and requires $\eta>0.65$. These results agree with our theorems in Sec.~\ref{sec:hierarchy}, $\rm{CR}\geq \rm{LR}=\rm{CM}\geq \rm{HZ}$. Besides, as the non-Gaussian state here is generated by single-photon subtraction to a Gaussian entangled state, it preserves Gaussian correlations that are also detected by first-order quadature witnesses.

Notably, a larger detection range can be obtained by introducing higher-order quadratures ($k=l=2$).
We show in Fig.~\ref{fig3} (c) that non-Gaussian steering is witnessed when $\eta>0.58$ for CR-type, $\eta>0.59$ for LR- and CM-type, and $\eta>0.62$ for HZ-type criterion, respectively. All of the required channel efficiencies become lower in comparison with the first-order quadratures case.

\subsection{Non-Gaussian mixtures of Gaussian states}
\begin{figure}[tb]
    \centering
    \includegraphics[width=8.5cm]{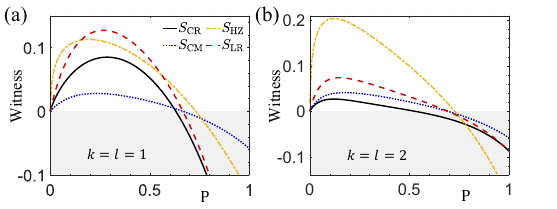}
    \caption{
	 Evolution of the different types of EPR steering criteria introduced in Sec.~\ref{sec:criteria} as function of the probability $P$ for a non-Gaussian state generated by mixing a TMSV and a SMSV with weights $P$ and $1-P$, respectively. Alice measures first-order quadratures ($k=l=1$) in (a), and second-order quadratures ($k=l=2$) in (b). The squeezing parameters of the SMSV states are set as $r_A=r_B=0.5$, and that of the TMSV state is $r=0.5$.} 
    \label{fig4}
\end{figure}

Finally, we discuss another typical example. As Gaussian states do not form a convex set, non-Gaussian states can be formed from convex mixtures of Gaussian states~\cite{PRXQuantum.2.030204}.
Here we consider a non-Gaussian state constructed by mixing a TMSV state with a single-mode squeezing state (SMSV), which reads as $ \hat{\rho}_s=P|\psi_{\mathrm{TMSV}}\rangle\langle\psi_{\mathrm{TMSV}}|_{AB}
    +(1-P)|\psi_{\mathrm{SMSV}}\rangle\langle\psi_{\mathrm{SMSV}}|_{AB},$
where $P$ represents the probability of measuring the TMSV state, and
$|\psi_{\mathrm{SMSV}}\rangle_{AB}=|\psi_{\mathrm{SMSV}}\rangle_{A}\otimes|\psi_{\mathrm{SMSV}}\rangle_{B}$ with
$
	|\psi_{\mathrm{SMSV}}\rangle_{A(B)} \propto \sum^{\infty}_{m=0}\tanh^{m}r_{A(B)}
	\sqrt{(2m)!} {(2^m m!)^{-1}} |2m\rangle_{A(B)}
$.
By applying a local canonical transformation, the CM can be expressed in a standard form with $c_1\neq-c_2$, where $\rm{LR}=\rm{CM}>\rm{HZ}$ is expected from Theorem 1.

We show in Figs.~\ref{fig4} (a) and (b) the evolution of the four non-Gaussian steering criteria with respect to the probability $P$ of the TMSV state for first-order quadratures ($k=l=1$) and second-order quadratures ($k=l=2$), respectively. Fig.~\ref{fig4}(a) shows that CM-type (blue dotted line) and LR-type (orange dashed line) non-Gaussian steering criteria have the same witnessing range $(P>0.666)$, while the HZ-type criterion (yellow dash-dot line) requires $P>0.730$. As expected, the CR-type criterion is always broader than others ($P>0.631$). Panel Fig.~\ref{fig4} (b) shows non-Gaussian steering detected by higher-order quadratures ($k=l=2$). By increasing the order of measured quadratures, steering can be correspondingly witnessed in a larger range of the TMSV-proportion, that is, $P>0.693$ for LR- and CM-type, $P>0.727$ for HZ-type criterion, and $P>0.528$ for CR-type, respectively. It is clear that in both cases the results are consistent with our theorems that state that the ability of the four criteria to detect non-Gaussian steering follow the hierarchy $\rm{CR}\geq \rm{LR}=\rm{CM}\geq \rm{HZ}$.

\section{Conclusion}

In conclusion, we proposed an efficient framework to measure this ``spooky" action at a distance based on high-order quadratures for arbitrary non-Gaussian systems. Other previously known steering criteria were generalized and  systematically categorized into a hierarchy. Our findings indicate that the CR-type criterion, which is based on conditional distributions, is the most effective in detecting non-Gaussian steering. In contrast, the CM-, LR-, and HZ-type criteria are weaker in this regard, with a relation of $ \rm{CR} \geq \rm{LR} = \rm{CM} \geq \rm{HZ}$. Only when the covariance matrix of the high-order quadratures can be expressed in the standard form with $c_1 = -c_2$, the CM-, LR-, and HZ-type criteria are all equivalent in their ability to detect non-Gaussian steering. We have verified our findings using three main experimentally realized non-Gaussian states, including three-photon SPDC states, photon-subtracted Gaussian states, and non-Gaussian mixtures of Gaussian states. Notably, the methods proposed here are general and not limited on any specific physical platforms, hence our analysis is trailblazing in exploring quantum correlations existing in higher-order observables for various systems, such as quantum optics~\cite{np11.713,lsh-prl}, superconducting circuits~\cite{spdcprx2020, wang2016schrodinger,wang2022flying}, trapped ions~\cite{ion-ref}, and cavity optomechanics~\cite{opomech2009,opomech2016}. Furthermore, it's potential to extend these conditions to multipartite systems for characterizing more complicated correlations~\cite{tian2022praaplied}. Last but not least, we also offered a scheme to experimentally produce the high-fidelity tunable multi-component Schr\"{o}dinger cat state, which is an essential resource for cutting-edge quantum computing. Comparing with other existing approaches ~\cite{cat2006,cat2007,cat2013}, high-order quadrature measurements allow for continuous tuning the size of the cat states, which holds significant benefits and applications for quantum information processing.

\appendix
\section{Derivation of Equation \eqref{wlr}} \label{appA}

We firstly analyze the elements of the standard-form CM of Eq. \eqref{cv}. By inserting the higher-order quadratures into the CM, we can obtain the following relations:
With $V_{11}=V_{22}=n$, we get $\langle \hat{A}^2\rangle + \langle \hat{A}^{\dagger 2}\rangle=0.$ And with $V_{12}=V_{21}=0$, it is directly checked that $\langle\hat{A}^2\rangle - \langle \hat{A}^{\dagger 2}\rangle=0$. Thus, we obtain $\langle \hat{A}^2\rangle=0$, and $\langle \hat{B}^2\rangle=0$ is similarly obtained. Besides, as $V_{23}=V_{14}=0$ we have $\langle \hat{A}\hat{B}-\hat{A}^\dagger \hat{B}^\dagger\rangle=0$ and $\langle \hat{A}\hat{B}^\dagger-\hat{A}^\dagger \hat{B}\rangle=0$, which indicates that $\langle AB\rangle$ and $\langle{AB^\dagger}\rangle$ are both real.
Then, we can express the matrix elements of the standard-form CM in terms of expectation values of the operators $\hat{A}$ and $\hat{B}$ as
\begin{equation}\label{eq:CM-parameters}
\begin{aligned}
         n&=\frac 14\langle \hat{A}\hat{A}^\dagger+\hat{A}^\dagger \hat{A}\rangle,~~~
         m=\frac 14\langle \hat{B}\hat{B}^\dagger+\hat{B}^\dagger \hat{B}\rangle,
         \\
         c_1&=\frac 12\langle \hat{A}\hat{B}^{\dagger} + \hat{A}\hat{B}\rangle,~~~~~~ 
         c_2=\frac 12\langle \hat{A}\hat{B}^{\dagger} - \hat{A}\hat{B}\rangle.
\end{aligned}
\end{equation}
Using the standard-form CM of Eq. \eqref{cv}, the matrix $M^{kl}_{B|A}$ in Eq.~\eqref{eqcm} can be written as 
\begin{equation}\label{Mba}
    M^{kl}_{B|A}=\left(\begin{array}{cccc}
    n & 0 & c_{1} & 0 \\
    0 & n & 0 & c_{2} \\
    c_{1} & 0 & m & iQ_{B} \\
    0 & c_{2} & -iQ_{B} & m
\end{array}\right).
\end{equation}
With the above derivation, the LR-steering criterion can be expanded with the elements of the CM, 
where the steerability from Alice to Bob is equivalent to
\begin{equation}\label{lr}
    [m+2c_1g_x+g_{x}^2 n][m-2c_2g_p+g_{p}^2 n]<Q_{B}^2.
\end{equation}
The optimal gains are 
 \begin{equation}\label{gop}
    g_x=-\frac{c_1}{n},\quad g_p=\frac{c_2}{n}.
\end{equation} 
Applying these gains in Eq. \eqref{lr} we obtain Eq. \eqref{wlr}.

\begin{acknowledgments}
{\it Acknowledgements.--}This work is supported by the National Natural Science Foundation of China (Grants No.~11975026, No. 12125402, No.~12004011, and No.~12147148), Beijing Natural Science Foundation (Grant No.~Z190005), the Key R\&D Program of Guangdong Province (Grant No. 2018B030329001), and the Innovation Program for Quantum Science and Technology (Grant No. 2021ZD0301500). F.-X.S. acknowledges the China Postdoctoral Science Foundation (Grant No.~2020M680186). K.B. recognizes support by the Agence Nationale de la Recherche through Project TRIQUI (No. ANR 17-CE24-0041).
\end{acknowledgments}



\begin{thebibliography}{75}%
\makeatletter
\providecommand \@ifxundefined [1]{%
 \@ifx{#1\undefined}
}%
\providecommand \@ifnum [1]{%
 \ifnum #1\expandafter \@firstoftwo
 \else \expandafter \@secondoftwo
 \fi
}%
\providecommand \@ifx [1]{%
 \ifx #1\expandafter \@firstoftwo
 \else \expandafter \@secondoftwo
 \fi
}%
\providecommand \natexlab [1]{#1}%
\providecommand \enquote  [1]{``#1''}%
\providecommand \bibnamefont  [1]{#1}%
\providecommand \bibfnamefont [1]{#1}%
\providecommand \citenamefont [1]{#1}%
\providecommand \href@noop [0]{\@secondoftwo}%
\providecommand \href [0]{\begingroup \@sanitize@url \@href}%
\providecommand \@href[1]{\@@startlink{#1}\@@href}%
\providecommand \@@href[1]{\endgroup#1\@@endlink}%
\providecommand \@sanitize@url [0]{\catcode `\\12\catcode `\$12\catcode
  `\&12\catcode `\#12\catcode `\^12\catcode `\_12\catcode `\%12\relax}%
\providecommand \@@startlink[1]{}%
\providecommand \@@endlink[0]{}%
\providecommand \url  [0]{\begingroup\@sanitize@url \@url }%
\providecommand \@url [1]{\endgroup\@href {#1}{\urlprefix }}%
\providecommand \urlprefix  [0]{URL }%
\providecommand \Eprint [0]{\href }%
\providecommand \doibase [0]{https://doi.org/}%
\providecommand \selectlanguage [0]{\@gobble}%
\providecommand \bibinfo  [0]{\@secondoftwo}%
\providecommand \bibfield  [0]{\@secondoftwo}%
\providecommand \translation [1]{[#1]}%
\providecommand \BibitemOpen [0]{}%
\providecommand \bibitemStop [0]{}%
\providecommand \bibitemNoStop [0]{.\EOS\space}%
\providecommand \EOS [0]{\spacefactor3000\relax}%
\providecommand \BibitemShut  [1]{\csname bibitem#1\endcsname}%
\let\auto@bib@innerbib\@empty
\bibitem [{\citenamefont {Einstein}\ \emph {et~al.}(1935)\citenamefont
  {Einstein}, \citenamefont {Podolsky},\ and\ \citenamefont
  {Rosen}}]{eprparadox1935}%
  \BibitemOpen
  \bibfield  {author} {\bibinfo {author} {\bibfnamefont {A.}~\bibnamefont
  {Einstein}}, \bibinfo {author} {\bibfnamefont {B.}~\bibnamefont {Podolsky}},\
  and\ \bibinfo {author} {\bibfnamefont {N.}~\bibnamefont {Rosen}},\ }\bibfield
   {title} {\bibinfo {title} {Can quantum-mechanical description of physical
  reality be considered complete?},\ }\href
  {https://doi.org/10.1103/PhysRev.47.777} {\bibfield  {journal} {\bibinfo
  {journal} {Phys. Rev.}\ }\textbf {\bibinfo {volume} {47}},\ \bibinfo {pages}
  {777} (\bibinfo {year} {1935})}\BibitemShut {NoStop}%
\bibitem [{\citenamefont {Schrödinger}(1935)}]{schrodinger1935}%
  \BibitemOpen
  \bibfield  {author} {\bibinfo {author} {\bibfnamefont {E.}~\bibnamefont
  {Schrödinger}},\ }\bibfield  {title} {\bibinfo {title} {Discussion of
  probability relations between separated systems},\ }\href
  {https://doi.org/10.1017/S0305004100013554} {\bibfield  {journal} {\bibinfo
  {journal} {Proc. Camb. Philos. Soc.}\ }\textbf {\bibinfo {volume} {31}},\
  \bibinfo {pages} {555–563} (\bibinfo {year} {1935})}\BibitemShut {NoStop}%
\bibitem [{\citenamefont {Wiseman}\ \emph {et~al.}(2007)\citenamefont
  {Wiseman}, \citenamefont {Jones},\ and\ \citenamefont
  {Doherty}}]{wiseman2007}%
  \BibitemOpen
  \bibfield  {author} {\bibinfo {author} {\bibfnamefont {H.~M.}\ \bibnamefont
  {Wiseman}}, \bibinfo {author} {\bibfnamefont {S.~J.}\ \bibnamefont {Jones}},\
  and\ \bibinfo {author} {\bibfnamefont {A.~C.}\ \bibnamefont {Doherty}},\
  }\bibfield  {title} {\bibinfo {title} {Steering, entanglement, nonlocality,
  and the {E}instein-{P}odolsky-{R}osen paradox},\ }\href
  {https://doi.org/10.1103/PhysRevLett.98.140402} {\bibfield  {journal}
  {\bibinfo  {journal} {Phys. Rev. Lett.}\ }\textbf {\bibinfo {volume} {98}},\
  \bibinfo {pages} {140402} (\bibinfo {year} {2007})}\BibitemShut {NoStop}%
\bibitem [{\citenamefont {Reid}\ \emph {et~al.}(2009)\citenamefont {Reid},
  \citenamefont {Drummond}, \citenamefont {Bowen}, \citenamefont {Cavalcanti},
  \citenamefont {Lam}, \citenamefont {Bachor}, \citenamefont {Andersen},\ and\
  \citenamefont {Leuchs}}]{rmp-2009}%
  \BibitemOpen
  \bibfield  {author} {\bibinfo {author} {\bibfnamefont {M.~D.}\ \bibnamefont
  {Reid}}, \bibinfo {author} {\bibfnamefont {P.~D.}\ \bibnamefont {Drummond}},
  \bibinfo {author} {\bibfnamefont {W.~P.}\ \bibnamefont {Bowen}}, \bibinfo
  {author} {\bibfnamefont {E.~G.}\ \bibnamefont {Cavalcanti}}, \bibinfo
  {author} {\bibfnamefont {P.~K.}\ \bibnamefont {Lam}}, \bibinfo {author}
  {\bibfnamefont {H.~A.}\ \bibnamefont {Bachor}}, \bibinfo {author}
  {\bibfnamefont {U.~L.}\ \bibnamefont {Andersen}},\ and\ \bibinfo {author}
  {\bibfnamefont {G.}~\bibnamefont {Leuchs}},\ }\bibfield  {title} {\bibinfo
  {title} {Colloquium: The {E}instein-{P}odolsky-{R}osen paradox: From concepts
  to applications},\ }\href {https://doi.org/10.1103/RevModPhys.81.1727}
  {\bibfield  {journal} {\bibinfo  {journal} {Rev. Mod. Phys.}\ }\textbf
  {\bibinfo {volume} {81}},\ \bibinfo {pages} {1727} (\bibinfo {year}
  {2009})}\BibitemShut {NoStop}%
\bibitem [{\citenamefont {Cavalcanti}\ and\ \citenamefont
  {Skrzypczyk}(2016)}]{review-steering-2016}%
  \BibitemOpen
  \bibfield  {author} {\bibinfo {author} {\bibfnamefont {D.}~\bibnamefont
  {Cavalcanti}}\ and\ \bibinfo {author} {\bibfnamefont {P.}~\bibnamefont
  {Skrzypczyk}},\ }\bibfield  {title} {\bibinfo {title} {Quantum steering: a
  review with focus on semidefinite programming},\ }\href
  {https://doi.org/10.1088/1361-6633/80/2/024001} {\bibfield  {journal}
  {\bibinfo  {journal} {Rep. Prog. Phys.}\ }\textbf {\bibinfo {volume} {80}},\
  \bibinfo {pages} {024001} (\bibinfo {year} {2016})}\BibitemShut {NoStop}%
\bibitem [{\citenamefont {Uola}\ \emph {et~al.}(2020)\citenamefont {Uola},
  \citenamefont {Costa}, \citenamefont {Nguyen},\ and\ \citenamefont
  {G\"uhne}}]{review-steering-rmp2020}%
  \BibitemOpen
  \bibfield  {author} {\bibinfo {author} {\bibfnamefont {R.}~\bibnamefont
  {Uola}}, \bibinfo {author} {\bibfnamefont {A.~C.~S.}\ \bibnamefont {Costa}},
  \bibinfo {author} {\bibfnamefont {H.~C.}\ \bibnamefont {Nguyen}},\ and\
  \bibinfo {author} {\bibfnamefont {O.}~\bibnamefont {G\"uhne}},\ }\bibfield
  {title} {\bibinfo {title} {Quantum steering},\ }\href
  {https://doi.org/10.1103/RevModPhys.92.015001} {\bibfield  {journal}
  {\bibinfo  {journal} {Rev. Mod. Phys.}\ }\textbf {\bibinfo {volume} {92}},\
  \bibinfo {pages} {015001} (\bibinfo {year} {2020})}\BibitemShut {NoStop}%
\bibitem [{\citenamefont {Xiang}\ \emph {et~al.}(2022)\citenamefont {Xiang},
  \citenamefont {Cheng}, \citenamefont {Gong}, \citenamefont {Ficek},\ and\
  \citenamefont {He}}]{xy-prxquantum2022}%
  \BibitemOpen
  \bibfield  {author} {\bibinfo {author} {\bibfnamefont {Y.}~\bibnamefont
  {Xiang}}, \bibinfo {author} {\bibfnamefont {S.}~\bibnamefont {Cheng}},
  \bibinfo {author} {\bibfnamefont {Q.}~\bibnamefont {Gong}}, \bibinfo {author}
  {\bibfnamefont {Z.}~\bibnamefont {Ficek}},\ and\ \bibinfo {author}
  {\bibfnamefont {Q.}~\bibnamefont {He}},\ }\bibfield  {title} {\bibinfo
  {title} {Quantum steering: Practical challenges and future directions},\
  }\href {https://doi.org/10.1103/PRXQuantum.3.030102} {\bibfield  {journal}
  {\bibinfo  {journal} {PRX Quantum}\ }\textbf {\bibinfo {volume} {3}},\
  \bibinfo {pages} {030102} (\bibinfo {year} {2022})}\BibitemShut {NoStop}%
\bibitem [{\citenamefont {Lo}\ \emph {et~al.}(2014)\citenamefont {Lo},
  \citenamefont {Curty},\ and\ \citenamefont {Tamaki}}]{QKD2014np}%
  \BibitemOpen
  \bibfield  {author} {\bibinfo {author} {\bibfnamefont {H.-K.}\ \bibnamefont
  {Lo}}, \bibinfo {author} {\bibfnamefont {M.}~\bibnamefont {Curty}},\ and\
  \bibinfo {author} {\bibfnamefont {K.}~\bibnamefont {Tamaki}},\ }\bibfield
  {title} {\bibinfo {title} {Secure quantum key distribution},\ }\href
  {https://doi.org/10.1038/nphoton.2014.149} {\bibfield  {journal} {\bibinfo
  {journal} {Nat. Photon.}\ }\textbf {\bibinfo {volume} {8}},\ \bibinfo {pages}
  {595} (\bibinfo {year} {2014})}\BibitemShut {NoStop}%
\bibitem [{\citenamefont {Gehring}\ \emph {et~al.}(2015)\citenamefont
  {Gehring}, \citenamefont {Händchen}, \citenamefont {Duhme}, \citenamefont
  {Furrer}, \citenamefont {Franz}, \citenamefont {Pacher}, \citenamefont
  {Werner},\ and\ \citenamefont {Schnabel}}]{QKD2015nc}%
  \BibitemOpen
  \bibfield  {author} {\bibinfo {author} {\bibfnamefont {T.}~\bibnamefont
  {Gehring}}, \bibinfo {author} {\bibfnamefont {V.}~\bibnamefont {Händchen}},
  \bibinfo {author} {\bibfnamefont {J.}~\bibnamefont {Duhme}}, \bibinfo
  {author} {\bibfnamefont {F.}~\bibnamefont {Furrer}}, \bibinfo {author}
  {\bibfnamefont {T.}~\bibnamefont {Franz}}, \bibinfo {author} {\bibfnamefont
  {C.}~\bibnamefont {Pacher}}, \bibinfo {author} {\bibfnamefont {R.~F.}\
  \bibnamefont {Werner}},\ and\ \bibinfo {author} {\bibfnamefont
  {R.}~\bibnamefont {Schnabel}},\ }\bibfield  {title} {\bibinfo {title}
  {Implementation of continuous-variable quantum key distribution with
  composable and one-sided-device-independent security against coherent
  attacks},\ }\href {https://doi.org/10.1038/ncomms9795} {\bibfield  {journal}
  {\bibinfo  {journal} {Nat. Commun.}\ }\textbf {\bibinfo {volume} {6}},\
  \bibinfo {pages} {8795} (\bibinfo {year} {2015})}\BibitemShut {NoStop}%
\bibitem [{\citenamefont {Walk}\ \emph {et~al.}(2016)\citenamefont {Walk},
  \citenamefont {Hosseini}, \citenamefont {Geng}, \citenamefont {Thearle},
  \citenamefont {Haw}, \citenamefont {Armstrong}, \citenamefont {Assad},
  \citenamefont {Janousek}, \citenamefont {Ralph}, \citenamefont {Symul},
  \citenamefont {Wiseman},\ and\ \citenamefont {Lam}}]{QKD-OPTICA}%
  \BibitemOpen
  \bibfield  {author} {\bibinfo {author} {\bibfnamefont {N.}~\bibnamefont
  {Walk}}, \bibinfo {author} {\bibfnamefont {S.}~\bibnamefont {Hosseini}},
  \bibinfo {author} {\bibfnamefont {J.}~\bibnamefont {Geng}}, \bibinfo {author}
  {\bibfnamefont {O.}~\bibnamefont {Thearle}}, \bibinfo {author} {\bibfnamefont
  {J.~Y.}\ \bibnamefont {Haw}}, \bibinfo {author} {\bibfnamefont
  {S.}~\bibnamefont {Armstrong}}, \bibinfo {author} {\bibfnamefont {S.~M.}\
  \bibnamefont {Assad}}, \bibinfo {author} {\bibfnamefont {J.}~\bibnamefont
  {Janousek}}, \bibinfo {author} {\bibfnamefont {T.~C.}\ \bibnamefont {Ralph}},
  \bibinfo {author} {\bibfnamefont {T.}~\bibnamefont {Symul}}, \bibinfo
  {author} {\bibfnamefont {H.~M.}\ \bibnamefont {Wiseman}},\ and\ \bibinfo
  {author} {\bibfnamefont {P.~K.}\ \bibnamefont {Lam}},\ }\bibfield  {title}
  {\bibinfo {title} {Experimental demonstration of {G}aussian protocols for
  one-sided device-independent quantum key distribution},\ }\href
  {https://doi.org/10.1364/OPTICA.3.000634} {\bibfield  {journal} {\bibinfo
  {journal} {Optica}\ }\textbf {\bibinfo {volume} {3}},\ \bibinfo {pages} {634}
  (\bibinfo {year} {2016})}\BibitemShut {NoStop}%
\bibitem [{\citenamefont {Armstrong}\ \emph {et~al.}(2015)\citenamefont
  {Armstrong}, \citenamefont {Wang}, \citenamefont {Teh}, \citenamefont {Gong},
  \citenamefont {He}, \citenamefont {Janousek}, \citenamefont {Bachor},
  \citenamefont {Reid},\ and\ \citenamefont {Lam}}]{he2015np}%
  \BibitemOpen
  \bibfield  {author} {\bibinfo {author} {\bibfnamefont {S.}~\bibnamefont
  {Armstrong}}, \bibinfo {author} {\bibfnamefont {M.}~\bibnamefont {Wang}},
  \bibinfo {author} {\bibfnamefont {R.~Y.}\ \bibnamefont {Teh}}, \bibinfo
  {author} {\bibfnamefont {Q.}~\bibnamefont {Gong}}, \bibinfo {author}
  {\bibfnamefont {Q.}~\bibnamefont {He}}, \bibinfo {author} {\bibfnamefont
  {J.}~\bibnamefont {Janousek}}, \bibinfo {author} {\bibfnamefont {H.-A.}\
  \bibnamefont {Bachor}}, \bibinfo {author} {\bibfnamefont {M.~D.}\
  \bibnamefont {Reid}},\ and\ \bibinfo {author} {\bibfnamefont {P.~K.}\
  \bibnamefont {Lam}},\ }\bibfield  {title} {\bibinfo {title} {Multipartite
  {E}instein–{P}odolsky–{R}osen steering and genuine tripartite
  entanglement with optical networks},\ }\href
  {https://doi.org/10.1038/nphys3202} {\bibfield  {journal} {\bibinfo
  {journal} {Nat. Phys.}\ }\textbf {\bibinfo {volume} {11}},\ \bibinfo {pages}
  {167} (\bibinfo {year} {2015})}\BibitemShut {NoStop}%
\bibitem [{\citenamefont {Kogias}\ \emph {et~al.}(2017)\citenamefont {Kogias},
  \citenamefont {Xiang}, \citenamefont {He},\ and\ \citenamefont
  {Adesso}}]{secretsharingpra2017}%
  \BibitemOpen
  \bibfield  {author} {\bibinfo {author} {\bibfnamefont {I.}~\bibnamefont
  {Kogias}}, \bibinfo {author} {\bibfnamefont {Y.}~\bibnamefont {Xiang}},
  \bibinfo {author} {\bibfnamefont {Q.}~\bibnamefont {He}},\ and\ \bibinfo
  {author} {\bibfnamefont {G.}~\bibnamefont {Adesso}},\ }\bibfield  {title}
  {\bibinfo {title} {Unconditional security of entanglement-based
  continuous-variable quantum secret sharing},\ }\href
  {https://doi.org/10.1103/PhysRevA.95.012315} {\bibfield  {journal} {\bibinfo
  {journal} {Phys. Rev. A}\ }\textbf {\bibinfo {volume} {95}},\ \bibinfo
  {pages} {012315} (\bibinfo {year} {2017})}\BibitemShut {NoStop}%
\bibitem [{\citenamefont {Passaro}\ \emph {et~al.}(2015)\citenamefont
  {Passaro}, \citenamefont {Cavalcanti}, \citenamefont {Skrzypczyk},\ and\
  \citenamefont {Acín}}]{randomness-njp}%
  \BibitemOpen
  \bibfield  {author} {\bibinfo {author} {\bibfnamefont {E.}~\bibnamefont
  {Passaro}}, \bibinfo {author} {\bibfnamefont {D.}~\bibnamefont {Cavalcanti}},
  \bibinfo {author} {\bibfnamefont {P.}~\bibnamefont {Skrzypczyk}},\ and\
  \bibinfo {author} {\bibfnamefont {A.}~\bibnamefont {Acín}},\ }\bibfield
  {title} {\bibinfo {title} {Optimal randomness certification in the quantum
  steering and prepare-and-measure scenarios},\ }\href
  {https://doi.org/10.1088/1367-2630/17/11/113010} {\bibfield  {journal}
  {\bibinfo  {journal} {New J. Phys.}\ }\textbf {\bibinfo {volume} {17}},\
  \bibinfo {pages} {113010} (\bibinfo {year} {2015})}\BibitemShut {NoStop}%
\bibitem [{\citenamefont {Skrzypczyk}\ and\ \citenamefont
  {Cavalcanti}(2018)}]{randomness-prl2018}%
  \BibitemOpen
  \bibfield  {author} {\bibinfo {author} {\bibfnamefont {P.}~\bibnamefont
  {Skrzypczyk}}\ and\ \bibinfo {author} {\bibfnamefont {D.}~\bibnamefont
  {Cavalcanti}},\ }\bibfield  {title} {\bibinfo {title} {Maximal randomness
  generation from steering inequality violations using qudits},\ }\href
  {https://doi.org/10.1103/PhysRevLett.120.260401} {\bibfield  {journal}
  {\bibinfo  {journal} {Phys. Rev. Lett.}\ }\textbf {\bibinfo {volume} {120}},\
  \bibinfo {pages} {260401} (\bibinfo {year} {2018})}\BibitemShut {NoStop}%
\bibitem [{\citenamefont {Guo}\ \emph {et~al.}(2019)\citenamefont {Guo},
  \citenamefont {Cheng}, \citenamefont {Hu}, \citenamefont {Liu}, \citenamefont
  {Huang}, \citenamefont {Huang}, \citenamefont {Li}, \citenamefont {Guo},\
  and\ \citenamefont {Cavalcanti}}]{randomness-prl2019}%
  \BibitemOpen
  \bibfield  {author} {\bibinfo {author} {\bibfnamefont {Y.}~\bibnamefont
  {Guo}}, \bibinfo {author} {\bibfnamefont {S.}~\bibnamefont {Cheng}}, \bibinfo
  {author} {\bibfnamefont {X.}~\bibnamefont {Hu}}, \bibinfo {author}
  {\bibfnamefont {B.-H.}\ \bibnamefont {Liu}}, \bibinfo {author} {\bibfnamefont
  {E.-M.}\ \bibnamefont {Huang}}, \bibinfo {author} {\bibfnamefont {Y.-F.}\
  \bibnamefont {Huang}}, \bibinfo {author} {\bibfnamefont {C.-F.}\ \bibnamefont
  {Li}}, \bibinfo {author} {\bibfnamefont {G.-C.}\ \bibnamefont {Guo}},\ and\
  \bibinfo {author} {\bibfnamefont {E.~G.}\ \bibnamefont {Cavalcanti}},\
  }\bibfield  {title} {\bibinfo {title} {Experimental
  measurement-device-independent quantum steering and randomness generation
  beyond qubits},\ }\href {https://doi.org/10.1103/PhysRevLett.123.170402}
  {\bibfield  {journal} {\bibinfo  {journal} {Phys. Rev. Lett.}\ }\textbf
  {\bibinfo {volume} {123}},\ \bibinfo {pages} {170402} (\bibinfo {year}
  {2019})}\BibitemShut {NoStop}%
\bibitem [{\citenamefont {Mari}\ and\ \citenamefont
  {Eisert}(2012)}]{PhysRevLett.109.230503}%
  \BibitemOpen
  \bibfield  {author} {\bibinfo {author} {\bibfnamefont {A.}~\bibnamefont
  {Mari}}\ and\ \bibinfo {author} {\bibfnamefont {J.}~\bibnamefont {Eisert}},\
  }\bibfield  {title} {\bibinfo {title} {Positive {W}igner functions render
  classical simulation of quantum computation efficient},\ }\href
  {https://doi.org/10.1103/PhysRevLett.109.230503} {\bibfield  {journal}
  {\bibinfo  {journal} {Phys. Rev. Lett.}\ }\textbf {\bibinfo {volume} {109}},\
  \bibinfo {pages} {230503} (\bibinfo {year} {2012})}\BibitemShut {NoStop}%
\bibitem [{\citenamefont {Karuseichyk}\ \emph {et~al.}(2022)\citenamefont
  {Karuseichyk}, \citenamefont {Sorelli}, \citenamefont {Walschaers},
  \citenamefont {Treps},\ and\ \citenamefont
  {Gessner}}]{PhysRevResearch.4.043010}%
  \BibitemOpen
  \bibfield  {author} {\bibinfo {author} {\bibfnamefont {I.}~\bibnamefont
  {Karuseichyk}}, \bibinfo {author} {\bibfnamefont {G.}~\bibnamefont
  {Sorelli}}, \bibinfo {author} {\bibfnamefont {M.}~\bibnamefont {Walschaers}},
  \bibinfo {author} {\bibfnamefont {N.}~\bibnamefont {Treps}},\ and\ \bibinfo
  {author} {\bibfnamefont {M.}~\bibnamefont {Gessner}},\ }\bibfield  {title}
  {\bibinfo {title} {Resolving mutually-coherent point sources of light with
  arbitrary statistics},\ }\href
  {https://doi.org/10.1103/PhysRevResearch.4.043010} {\bibfield  {journal}
  {\bibinfo  {journal} {Phys. Rev. Res.}\ }\textbf {\bibinfo {volume} {4}},\
  \bibinfo {pages} {043010} (\bibinfo {year} {2022})}\BibitemShut {NoStop}%
\bibitem [{\citenamefont {Andersen}\ \emph {et~al.}(2015)\citenamefont
  {Andersen}, \citenamefont {Neergaard-Nielsen}, \citenamefont {van Loock},\
  and\ \citenamefont {Furusawa}}]{np11.713}%
  \BibitemOpen
  \bibfield  {author} {\bibinfo {author} {\bibfnamefont {U.~L.}\ \bibnamefont
  {Andersen}}, \bibinfo {author} {\bibfnamefont {J.~S.}\ \bibnamefont
  {Neergaard-Nielsen}}, \bibinfo {author} {\bibfnamefont {P.}~\bibnamefont {van
  Loock}},\ and\ \bibinfo {author} {\bibfnamefont {A.}~\bibnamefont
  {Furusawa}},\ }\bibfield  {title} {\bibinfo {title} {Hybrid discrete- and
  continuous-variable quantum information},\ }\href
  {https://doi.org/10.1038/nphys3410} {\bibfield  {journal} {\bibinfo
  {journal} {Nat. Phys.}\ }\textbf {\bibinfo {volume} {11}},\ \bibinfo {pages}
  {713} (\bibinfo {year} {2015})}\BibitemShut {NoStop}%
\bibitem [{\citenamefont {Liu}\ \emph {et~al.}(2022)\citenamefont {Liu},
  \citenamefont {Han}, \citenamefont {Wang}, \citenamefont {Xiang},
  \citenamefont {Sun}, \citenamefont {Wang}, \citenamefont {Qin}, \citenamefont
  {Gong}, \citenamefont {Su},\ and\ \citenamefont {He}}]{lsh-prl}%
  \BibitemOpen
  \bibfield  {author} {\bibinfo {author} {\bibfnamefont {S.}~\bibnamefont
  {Liu}}, \bibinfo {author} {\bibfnamefont {D.}~\bibnamefont {Han}}, \bibinfo
  {author} {\bibfnamefont {N.}~\bibnamefont {Wang}}, \bibinfo {author}
  {\bibfnamefont {Y.}~\bibnamefont {Xiang}}, \bibinfo {author} {\bibfnamefont
  {F.}~\bibnamefont {Sun}}, \bibinfo {author} {\bibfnamefont {M.}~\bibnamefont
  {Wang}}, \bibinfo {author} {\bibfnamefont {Z.}~\bibnamefont {Qin}}, \bibinfo
  {author} {\bibfnamefont {Q.}~\bibnamefont {Gong}}, \bibinfo {author}
  {\bibfnamefont {X.}~\bibnamefont {Su}},\ and\ \bibinfo {author}
  {\bibfnamefont {Q.}~\bibnamefont {He}},\ }\bibfield  {title} {\bibinfo
  {title} {Experimental demonstration of remotely creating {W}igner negativity
  via quantum steering},\ }\href
  {https://doi.org/10.1103/PhysRevLett.128.200401} {\bibfield  {journal}
  {\bibinfo  {journal} {Phys. Rev. Lett.}\ }\textbf {\bibinfo {volume} {128}},\
  \bibinfo {pages} {200401} (\bibinfo {year} {2022})}\BibitemShut {NoStop}%
\bibitem [{\citenamefont {G\'omez}\ \emph {et~al.}(2015)\citenamefont
  {G\'omez}, \citenamefont {Ca\~nas}, \citenamefont {Acu\~na}, \citenamefont
  {Nogueira},\ and\ \citenamefont {Lima}}]{pra2015}%
  \BibitemOpen
  \bibfield  {author} {\bibinfo {author} {\bibfnamefont {E.~S.}\ \bibnamefont
  {G\'omez}}, \bibinfo {author} {\bibfnamefont {G.}~\bibnamefont {Ca\~nas}},
  \bibinfo {author} {\bibfnamefont {E.}~\bibnamefont {Acu\~na}}, \bibinfo
  {author} {\bibfnamefont {W.~A.~T.}\ \bibnamefont {Nogueira}},\ and\ \bibinfo
  {author} {\bibfnamefont {G.}~\bibnamefont {Lima}},\ }\bibfield  {title}
  {\bibinfo {title} {Non-{G}aussian-state generation certified using the
  {E}instein-{P}odolsky-{R}osen-steering inequality},\ }\href
  {https://doi.org/10.1103/PhysRevA.91.013801} {\bibfield  {journal} {\bibinfo
  {journal} {Phys. Rev. A}\ }\textbf {\bibinfo {volume} {91}},\ \bibinfo
  {pages} {013801} (\bibinfo {year} {2015})}\BibitemShut {NoStop}%
\bibitem [{\citenamefont {Douady}\ and\ \citenamefont
  {Boulanger}(2004)}]{spdc2004}%
  \BibitemOpen
  \bibfield  {author} {\bibinfo {author} {\bibfnamefont {J.}~\bibnamefont
  {Douady}}\ and\ \bibinfo {author} {\bibfnamefont {B.}~\bibnamefont
  {Boulanger}},\ }\bibfield  {title} {\bibinfo {title} {Experimental
  demonstration of a pure third-order optical parametric downconversion
  process},\ }\href {https://doi.org/10.1364/OL.29.002794} {\bibfield
  {journal} {\bibinfo  {journal} {Opt. Lett.}\ }\textbf {\bibinfo {volume}
  {29}},\ \bibinfo {pages} {2794} (\bibinfo {year} {2004})}\BibitemShut
  {NoStop}%
\bibitem [{\citenamefont {Chang}\ \emph {et~al.}(2020)\citenamefont {Chang},
  \citenamefont {Sab\'{\i}n}, \citenamefont {Forn-D\'{\i}az}, \citenamefont
  {Quijandr\'{\i}a}, \citenamefont {Vadiraj}, \citenamefont {Nsanzineza},
  \citenamefont {Johansson},\ and\ \citenamefont {Wilson}}]{spdcprx2020}%
  \BibitemOpen
  \bibfield  {author} {\bibinfo {author} {\bibfnamefont {C.~W.~S.}\
  \bibnamefont {Chang}}, \bibinfo {author} {\bibfnamefont {C.}~\bibnamefont
  {Sab\'{\i}n}}, \bibinfo {author} {\bibfnamefont {P.}~\bibnamefont
  {Forn-D\'{\i}az}}, \bibinfo {author} {\bibfnamefont {F.}~\bibnamefont
  {Quijandr\'{\i}a}}, \bibinfo {author} {\bibfnamefont {A.~M.}\ \bibnamefont
  {Vadiraj}}, \bibinfo {author} {\bibfnamefont {I.}~\bibnamefont {Nsanzineza}},
  \bibinfo {author} {\bibfnamefont {G.}~\bibnamefont {Johansson}},\ and\
  \bibinfo {author} {\bibfnamefont {C.~M.}\ \bibnamefont {Wilson}},\ }\bibfield
   {title} {\bibinfo {title} {Observation of three-photon spontaneous
  parametric down-conversion in a superconducting parametric cavity},\ }\href
  {https://doi.org/10.1103/PhysRevX.10.011011} {\bibfield  {journal} {\bibinfo
  {journal} {Phys. Rev. X}\ }\textbf {\bibinfo {volume} {10}},\ \bibinfo
  {pages} {011011} (\bibinfo {year} {2020})}\BibitemShut {NoStop}%
\bibitem [{\citenamefont {Wang}\ \emph {et~al.}(2016)\citenamefont {Wang},
  \citenamefont {Gao}, \citenamefont {Reinhold}, \citenamefont {Heeres},
  \citenamefont {Ofek}, \citenamefont {Chou}, \citenamefont {Axline},
  \citenamefont {Reagor}, \citenamefont {Blumoff}, \citenamefont {Sliwa},
  \citenamefont {Frunzio}, \citenamefont {Girvin}, \citenamefont {Jiang},
  \citenamefont {Mirrahimi}, \citenamefont {Devoret},\ and\ \citenamefont
  {Schoelkopf}}]{wang2016schrodinger}%
  \BibitemOpen
  \bibfield  {author} {\bibinfo {author} {\bibfnamefont {C.}~\bibnamefont
  {Wang}}, \bibinfo {author} {\bibfnamefont {Y.~Y.}\ \bibnamefont {Gao}},
  \bibinfo {author} {\bibfnamefont {P.}~\bibnamefont {Reinhold}}, \bibinfo
  {author} {\bibfnamefont {R.~W.}\ \bibnamefont {Heeres}}, \bibinfo {author}
  {\bibfnamefont {N.}~\bibnamefont {Ofek}}, \bibinfo {author} {\bibfnamefont
  {K.}~\bibnamefont {Chou}}, \bibinfo {author} {\bibfnamefont {C.}~\bibnamefont
  {Axline}}, \bibinfo {author} {\bibfnamefont {M.}~\bibnamefont {Reagor}},
  \bibinfo {author} {\bibfnamefont {J.}~\bibnamefont {Blumoff}}, \bibinfo
  {author} {\bibfnamefont {K.~M.}\ \bibnamefont {Sliwa}}, \bibinfo {author}
  {\bibfnamefont {L.}~\bibnamefont {Frunzio}}, \bibinfo {author} {\bibfnamefont
  {S.~M.}\ \bibnamefont {Girvin}}, \bibinfo {author} {\bibfnamefont
  {L.}~\bibnamefont {Jiang}}, \bibinfo {author} {\bibfnamefont
  {M.}~\bibnamefont {Mirrahimi}}, \bibinfo {author} {\bibfnamefont {M.~H.}\
  \bibnamefont {Devoret}},\ and\ \bibinfo {author} {\bibfnamefont {R.~J.}\
  \bibnamefont {Schoelkopf}},\ }\bibfield  {title} {\bibinfo {title} {A
  schr{\"o}dinger cat living in two boxes},\ }\href
  {https://doi.org/10.1126/science.aaf2941} {\bibfield  {journal} {\bibinfo
  {journal} {Science}\ }\textbf {\bibinfo {volume} {352}},\ \bibinfo {pages}
  {1087} (\bibinfo {year} {2016})}\BibitemShut {NoStop}%
\bibitem [{\citenamefont {Wang}\ \emph {et~al.}(2022)\citenamefont {Wang},
  \citenamefont {Bao}, \citenamefont {Wu}, \citenamefont {Li}, \citenamefont
  {Cai}, \citenamefont {Wang}, \citenamefont {Ma}, \citenamefont {Cai},
  \citenamefont {Han}, \citenamefont {Wang}, \citenamefont {Song},
  \citenamefont {Sun}, \citenamefont {Zhang},\ and\ \citenamefont
  {Duan}}]{wang2022flying}%
  \BibitemOpen
  \bibfield  {author} {\bibinfo {author} {\bibfnamefont {Z.}~\bibnamefont
  {Wang}}, \bibinfo {author} {\bibfnamefont {Z.}~\bibnamefont {Bao}}, \bibinfo
  {author} {\bibfnamefont {Y.}~\bibnamefont {Wu}}, \bibinfo {author}
  {\bibfnamefont {Y.}~\bibnamefont {Li}}, \bibinfo {author} {\bibfnamefont
  {W.}~\bibnamefont {Cai}}, \bibinfo {author} {\bibfnamefont {W.}~\bibnamefont
  {Wang}}, \bibinfo {author} {\bibfnamefont {Y.}~\bibnamefont {Ma}}, \bibinfo
  {author} {\bibfnamefont {T.}~\bibnamefont {Cai}}, \bibinfo {author}
  {\bibfnamefont {X.}~\bibnamefont {Han}}, \bibinfo {author} {\bibfnamefont
  {J.}~\bibnamefont {Wang}}, \bibinfo {author} {\bibfnamefont {Y.}~\bibnamefont
  {Song}}, \bibinfo {author} {\bibfnamefont {L.}~\bibnamefont {Sun}}, \bibinfo
  {author} {\bibfnamefont {H.}~\bibnamefont {Zhang}},\ and\ \bibinfo {author}
  {\bibfnamefont {L.}~\bibnamefont {Duan}},\ }\bibfield  {title} {\bibinfo
  {title} {A flying {S}chr\"{o}dinger's cat in multipartite entangled states},\
  }\href {https://doi.org/10.1126/sciadv.abn1778} {\bibfield  {journal}
  {\bibinfo  {journal} {Sci. Adv.}\ }\textbf {\bibinfo {volume} {8}},\ \bibinfo
  {pages} {eabn1778} (\bibinfo {year} {2022})}\BibitemShut {NoStop}%
\bibitem [{\citenamefont {Walschaers}(2021)}]{PRXQuantum.2.030204}%
  \BibitemOpen
  \bibfield  {author} {\bibinfo {author} {\bibfnamefont {M.}~\bibnamefont
  {Walschaers}},\ }\bibfield  {title} {\bibinfo {title} {Non-{G}aussian quantum
  states and where to find them},\ }\href
  {https://doi.org/10.1103/PRXQuantum.2.030204} {\bibfield  {journal} {\bibinfo
   {journal} {PRX Quantum}\ }\textbf {\bibinfo {volume} {2}},\ \bibinfo {pages}
  {030204} (\bibinfo {year} {2021})}\BibitemShut {NoStop}%
\bibitem [{\citenamefont {Eisert}\ \emph {et~al.}(2002)\citenamefont {Eisert},
  \citenamefont {Scheel},\ and\ \citenamefont {Plenio}}]{eisert2002distilling}%
  \BibitemOpen
  \bibfield  {author} {\bibinfo {author} {\bibfnamefont {J.}~\bibnamefont
  {Eisert}}, \bibinfo {author} {\bibfnamefont {S.}~\bibnamefont {Scheel}},\
  and\ \bibinfo {author} {\bibfnamefont {M.~B.}\ \bibnamefont {Plenio}},\
  }\bibfield  {title} {\bibinfo {title} {Distilling {G}aussian states with
  {G}aussian operations is impossible},\ }\href
  {https://doi.org/10.1103/PhysRevLett.89.137903} {\bibfield  {journal}
  {\bibinfo  {journal} {Phys. Rev. Lett.}\ }\textbf {\bibinfo {volume} {89}},\
  \bibinfo {pages} {137903} (\bibinfo {year} {2002})}\BibitemShut {NoStop}%
\bibitem [{\citenamefont {Fiur{\'a}{\v{s}}ek}(2002)}]{fiuravsek2002gaussian}%
  \BibitemOpen
  \bibfield  {author} {\bibinfo {author} {\bibfnamefont {J.}~\bibnamefont
  {Fiur{\'a}{\v{s}}ek}},\ }\bibfield  {title} {\bibinfo {title} {Gaussian
  transformations and distillation of entangled {G}aussian states},\ }\href
  {https://doi.org/10.1103/PhysRevLett.89.137904} {\bibfield  {journal}
  {\bibinfo  {journal} {Phys. Rev. Lett.}\ }\textbf {\bibinfo {volume} {89}},\
  \bibinfo {pages} {137904} (\bibinfo {year} {2002})}\BibitemShut {NoStop}%
\bibitem [{\citenamefont {Takahashi}\ \emph {et~al.}(2010)\citenamefont
  {Takahashi}, \citenamefont {Neergaard-Nielsen}, \citenamefont {Takeuchi},
  \citenamefont {Takeoka}, \citenamefont {Hayasaka}, \citenamefont {Furusawa},\
  and\ \citenamefont {Sasaki}}]{nphoton.2010.1}%
  \BibitemOpen
  \bibfield  {author} {\bibinfo {author} {\bibfnamefont {H.}~\bibnamefont
  {Takahashi}}, \bibinfo {author} {\bibfnamefont {J.~S.}\ \bibnamefont
  {Neergaard-Nielsen}}, \bibinfo {author} {\bibfnamefont {M.}~\bibnamefont
  {Takeuchi}}, \bibinfo {author} {\bibfnamefont {M.}~\bibnamefont {Takeoka}},
  \bibinfo {author} {\bibfnamefont {K.}~\bibnamefont {Hayasaka}}, \bibinfo
  {author} {\bibfnamefont {A.}~\bibnamefont {Furusawa}},\ and\ \bibinfo
  {author} {\bibfnamefont {M.}~\bibnamefont {Sasaki}},\ }\bibfield  {title}
  {\bibinfo {title} {Entanglement distillation from {G}aussian input states},\
  }\href {https://doi.org/10.1038/nphoton.2010.1} {\bibfield  {journal}
  {\bibinfo  {journal} {Nat. Photon.}\ }\textbf {\bibinfo {volume} {4}},\
  \bibinfo {pages} {178} (\bibinfo {year} {2010})}\BibitemShut {NoStop}%
\bibitem [{\citenamefont {Ji}\ \emph {et~al.}(2016)\citenamefont {Ji},
  \citenamefont {Lee}, \citenamefont {Park},\ and\ \citenamefont
  {Nha}}]{nha-scirep2016}%
  \BibitemOpen
  \bibfield  {author} {\bibinfo {author} {\bibfnamefont {S.-W.}\ \bibnamefont
  {Ji}}, \bibinfo {author} {\bibfnamefont {J.}~\bibnamefont {Lee}}, \bibinfo
  {author} {\bibfnamefont {J.}~\bibnamefont {Park}},\ and\ \bibinfo {author}
  {\bibfnamefont {H.}~\bibnamefont {Nha}},\ }\bibfield  {title} {\bibinfo
  {title} {Quantum steering of {G}aussian states via non-{G}aussian
  measurements},\ }\href {https://doi.org/10.1038/srep29729} {\bibfield
  {journal} {\bibinfo  {journal} {Sci. Rep.}\ }\textbf {\bibinfo {volume}
  {6}},\ \bibinfo {pages} {29729} (\bibinfo {year} {2016})}\BibitemShut
  {NoStop}%
\bibitem [{\citenamefont {Strobel}\ \emph {et~al.}(2014)\citenamefont
  {Strobel}, \citenamefont {Muessel}, \citenamefont {Linnemann}, \citenamefont
  {Zibold}, \citenamefont {Hume}, \citenamefont {Pezzè}, \citenamefont
  {Smerzi},\ and\ \citenamefont {Oberthaler}}]{science.345}%
  \BibitemOpen
  \bibfield  {author} {\bibinfo {author} {\bibfnamefont {H.}~\bibnamefont
  {Strobel}}, \bibinfo {author} {\bibfnamefont {W.}~\bibnamefont {Muessel}},
  \bibinfo {author} {\bibfnamefont {D.}~\bibnamefont {Linnemann}}, \bibinfo
  {author} {\bibfnamefont {T.}~\bibnamefont {Zibold}}, \bibinfo {author}
  {\bibfnamefont {D.~B.}\ \bibnamefont {Hume}}, \bibinfo {author}
  {\bibfnamefont {L.}~\bibnamefont {Pezzè}}, \bibinfo {author} {\bibfnamefont
  {A.}~\bibnamefont {Smerzi}},\ and\ \bibinfo {author} {\bibfnamefont {M.~K.}\
  \bibnamefont {Oberthaler}},\ }\bibfield  {title} {\bibinfo {title} {{F}isher
  information and entanglement of non-{G}aussian spin states},\ }\href
  {https://doi.org/10.1126/science.1250147} {\bibfield  {journal} {\bibinfo
  {journal} {Science}\ }\textbf {\bibinfo {volume} {345}},\ \bibinfo {pages}
  {424} (\bibinfo {year} {2014})}\BibitemShut {NoStop}%
\bibitem [{\citenamefont {Joo}\ \emph {et~al.}(2011)\citenamefont {Joo},
  \citenamefont {Munro},\ and\ \citenamefont
  {Spiller}}]{PhysRevLett.107.083601}%
  \BibitemOpen
  \bibfield  {author} {\bibinfo {author} {\bibfnamefont {J.}~\bibnamefont
  {Joo}}, \bibinfo {author} {\bibfnamefont {W.~J.}\ \bibnamefont {Munro}},\
  and\ \bibinfo {author} {\bibfnamefont {T.~P.}\ \bibnamefont {Spiller}},\
  }\bibfield  {title} {\bibinfo {title} {Quantum metrology with entangled
  coherent states},\ }\href {https://doi.org/10.1103/PhysRevLett.107.083601}
  {\bibfield  {journal} {\bibinfo  {journal} {Phys. Rev. Lett.}\ }\textbf
  {\bibinfo {volume} {107}},\ \bibinfo {pages} {083601} (\bibinfo {year}
  {2011})}\BibitemShut {NoStop}%
\bibitem [{\citenamefont {Liu}\ \emph {et~al.}(2021)\citenamefont {Liu},
  \citenamefont {Tian}, \citenamefont {Liu}, \citenamefont {Dong},
  \citenamefont {Guo}, \citenamefont {He}, \citenamefont {Xu},\ and\
  \citenamefont {Li}}]{PhysRevApplied.16.064037}%
  \BibitemOpen
  \bibfield  {author} {\bibinfo {author} {\bibfnamefont {D.}~\bibnamefont
  {Liu}}, \bibinfo {author} {\bibfnamefont {M.}~\bibnamefont {Tian}}, \bibinfo
  {author} {\bibfnamefont {S.}~\bibnamefont {Liu}}, \bibinfo {author}
  {\bibfnamefont {X.}~\bibnamefont {Dong}}, \bibinfo {author} {\bibfnamefont
  {J.}~\bibnamefont {Guo}}, \bibinfo {author} {\bibfnamefont {Q.}~\bibnamefont
  {He}}, \bibinfo {author} {\bibfnamefont {H.}~\bibnamefont {Xu}},\ and\
  \bibinfo {author} {\bibfnamefont {Z.}~\bibnamefont {Li}},\ }\bibfield
  {title} {\bibinfo {title} {Ghost imaging with non-{G}aussian quantum light},\
  }\href {https://doi.org/10.1103/PhysRevApplied.16.064037} {\bibfield
  {journal} {\bibinfo  {journal} {Phys. Rev. Appl.}\ }\textbf {\bibinfo
  {volume} {16}},\ \bibinfo {pages} {064037} (\bibinfo {year}
  {2021})}\BibitemShut {NoStop}%
\bibitem [{\citenamefont {Ilya}\ \emph {et~al.}(2021)\citenamefont {Ilya},
  \citenamefont {Giacomo}, \citenamefont {Manuel}, \citenamefont {Mattia},\
  and\ \citenamefont {Nicolas}}]{mattia}%
  \BibitemOpen
  \bibfield  {author} {\bibinfo {author} {\bibfnamefont {K.}~\bibnamefont
  {Ilya}}, \bibinfo {author} {\bibfnamefont {S.}~\bibnamefont {Giacomo}},
  \bibinfo {author} {\bibfnamefont {G.}~\bibnamefont {Manuel}}, \bibinfo
  {author} {\bibfnamefont {W.}~\bibnamefont {Mattia}},\ and\ \bibinfo {author}
  {\bibfnamefont {T.}~\bibnamefont {Nicolas}},\ }\bibfield  {title} {\bibinfo
  {title} {Resolving mutually coherent bright point-sources},\ }\href
  {https://doi.org/10.48550/arXiv.2111.02233} {\bibfield  {journal} {\bibinfo
  {journal} {arXiv:2111.02233}\ } (\bibinfo {year} {2021})}\BibitemShut
  {NoStop}%
\bibitem [{\citenamefont {Chabaud}\ and\ \citenamefont
  {Walschaers}(2023)}]{PhysRevLett.130.090602}%
  \BibitemOpen
  \bibfield  {author} {\bibinfo {author} {\bibfnamefont {U.}~\bibnamefont
  {Chabaud}}\ and\ \bibinfo {author} {\bibfnamefont {M.}~\bibnamefont
  {Walschaers}},\ }\bibfield  {title} {\bibinfo {title} {Resources for bosonic
  quantum computational advantage},\ }\href
  {https://doi.org/10.1103/PhysRevLett.130.090602} {\bibfield  {journal}
  {\bibinfo  {journal} {Phys. Rev. Lett.}\ }\textbf {\bibinfo {volume} {130}},\
  \bibinfo {pages} {090602} (\bibinfo {year} {2023})}\BibitemShut {NoStop}%
\bibitem [{\citenamefont {Simon}(2000)}]{simon}%
  \BibitemOpen
  \bibfield  {author} {\bibinfo {author} {\bibfnamefont {R.}~\bibnamefont
  {Simon}},\ }\bibfield  {title} {\bibinfo {title} {Peres-{H}orodecki
  separability criterion for continuous variable systems},\ }\href
  {https://doi.org/10.1103/PhysRevLett.84.2726} {\bibfield  {journal} {\bibinfo
   {journal} {Phys. Rev. Lett.}\ }\textbf {\bibinfo {volume} {84}},\ \bibinfo
  {pages} {2726} (\bibinfo {year} {2000})}\BibitemShut {NoStop}%
\bibitem [{\citenamefont {van Loock}\ and\ \citenamefont
  {Furusawa}(2003)}]{van2003}%
  \BibitemOpen
  \bibfield  {author} {\bibinfo {author} {\bibfnamefont {P.}~\bibnamefont {van
  Loock}}\ and\ \bibinfo {author} {\bibfnamefont {A.}~\bibnamefont
  {Furusawa}},\ }\bibfield  {title} {\bibinfo {title} {Detecting genuine
  multipartite continuous-variable entanglement},\ }\href
  {https://doi.org/10.1103/PhysRevA.67.052315} {\bibfield  {journal} {\bibinfo
  {journal} {Phys. Rev. A}\ }\textbf {\bibinfo {volume} {67}},\ \bibinfo
  {pages} {052315} (\bibinfo {year} {2003})}\BibitemShut {NoStop}%
\bibitem [{\citenamefont {Adesso}\ and\ \citenamefont
  {Illuminati}(2005)}]{gerardo2005}%
  \BibitemOpen
  \bibfield  {author} {\bibinfo {author} {\bibfnamefont {G.}~\bibnamefont
  {Adesso}}\ and\ \bibinfo {author} {\bibfnamefont {F.}~\bibnamefont
  {Illuminati}},\ }\bibfield  {title} {\bibinfo {title} {Gaussian measures of
  entanglement versus negativities: Ordering of two-mode {G}aussian states},\
  }\href {https://doi.org/10.1103/PhysRevA.72.032334} {\bibfield  {journal}
  {\bibinfo  {journal} {Phys. Rev. A}\ }\textbf {\bibinfo {volume} {72}},\
  \bibinfo {pages} {032334} (\bibinfo {year} {2005})}\BibitemShut {NoStop}%
\bibitem [{\citenamefont {Hillery}\ and\ \citenamefont
  {Zubairy}(2006)}]{hz2006}%
  \BibitemOpen
  \bibfield  {author} {\bibinfo {author} {\bibfnamefont {M.}~\bibnamefont
  {Hillery}}\ and\ \bibinfo {author} {\bibfnamefont {M.~S.}\ \bibnamefont
  {Zubairy}},\ }\bibfield  {title} {\bibinfo {title} {Entanglement conditions
  for two-mode states},\ }\href {https://doi.org/10.1103/PhysRevLett.96.050503}
  {\bibfield  {journal} {\bibinfo  {journal} {Phys. Rev. Lett.}\ }\textbf
  {\bibinfo {volume} {96}},\ \bibinfo {pages} {050503} (\bibinfo {year}
  {2006})}\BibitemShut {NoStop}%
\bibitem [{\citenamefont {Teh}\ and\ \citenamefont {Reid}(2014)}]{reid2014pra}%
  \BibitemOpen
  \bibfield  {author} {\bibinfo {author} {\bibfnamefont {R.~Y.}\ \bibnamefont
  {Teh}}\ and\ \bibinfo {author} {\bibfnamefont {M.~D.}\ \bibnamefont {Reid}},\
  }\bibfield  {title} {\bibinfo {title} {Criteria for genuine $n$-partite
  continuous-variable entanglement and {E}instein-{P}odolsky-{R}osen
  steering},\ }\href {https://doi.org/10.1103/PhysRevA.90.062337} {\bibfield
  {journal} {\bibinfo  {journal} {Phys. Rev. A}\ }\textbf {\bibinfo {volume}
  {90}},\ \bibinfo {pages} {062337} (\bibinfo {year} {2014})}\BibitemShut
  {NoStop}%
\bibitem [{\citenamefont {He}\ and\ \citenamefont {Reid}(2013)}]{he2015prl}%
  \BibitemOpen
  \bibfield  {author} {\bibinfo {author} {\bibfnamefont {Q.~Y.}\ \bibnamefont
  {He}}\ and\ \bibinfo {author} {\bibfnamefont {M.~D.}\ \bibnamefont {Reid}},\
  }\bibfield  {title} {\bibinfo {title} {Genuine multipartite
  {E}instein-{P}odolsky-{R}osen steering},\ }\href
  {https://doi.org/10.1103/PhysRevLett.111.250403} {\bibfield  {journal}
  {\bibinfo  {journal} {Phys. Rev. Lett.}\ }\textbf {\bibinfo {volume} {111}},\
  \bibinfo {pages} {250403} (\bibinfo {year} {2013})}\BibitemShut {NoStop}%
\bibitem [{\citenamefont {Agust\'{\i}}\ \emph {et~al.}(2020)\citenamefont
  {Agust\'{\i}}, \citenamefont {Chang}, \citenamefont {Quijandr\'{\i}a},
  \citenamefont {Johansson}, \citenamefont {Wilson},\ and\ \citenamefont
  {Sab\'{\i}n}}]{hzspdc2020}%
  \BibitemOpen
  \bibfield  {author} {\bibinfo {author} {\bibfnamefont {A.}~\bibnamefont
  {Agust\'{\i}}}, \bibinfo {author} {\bibfnamefont {C.~W.~S.}\ \bibnamefont
  {Chang}}, \bibinfo {author} {\bibfnamefont {F.}~\bibnamefont
  {Quijandr\'{\i}a}}, \bibinfo {author} {\bibfnamefont {G.}~\bibnamefont
  {Johansson}}, \bibinfo {author} {\bibfnamefont {C.~M.}\ \bibnamefont
  {Wilson}},\ and\ \bibinfo {author} {\bibfnamefont {C.}~\bibnamefont
  {Sab\'{\i}n}},\ }\bibfield  {title} {\bibinfo {title} {Tripartite genuine
  non-{G}aussian entanglement in three-mode spontaneous parametric
  down-conversion},\ }\href {https://doi.org/10.1103/PhysRevLett.125.020502}
  {\bibfield  {journal} {\bibinfo  {journal} {Phys. Rev. Lett.}\ }\textbf
  {\bibinfo {volume} {125}},\ \bibinfo {pages} {020502} (\bibinfo {year}
  {2020})}\BibitemShut {NoStop}%
\bibitem [{\citenamefont {Shen}\ \emph {et~al.}(2015)\citenamefont {Shen},
  \citenamefont {Assad}, \citenamefont {Grosse}, \citenamefont {Li},
  \citenamefont {Reid},\ and\ \citenamefont {Lam}}]{reid2015}%
  \BibitemOpen
  \bibfield  {author} {\bibinfo {author} {\bibfnamefont {Y.}~\bibnamefont
  {Shen}}, \bibinfo {author} {\bibfnamefont {S.~M.}\ \bibnamefont {Assad}},
  \bibinfo {author} {\bibfnamefont {N.~B.}\ \bibnamefont {Grosse}}, \bibinfo
  {author} {\bibfnamefont {X.~Y.}\ \bibnamefont {Li}}, \bibinfo {author}
  {\bibfnamefont {M.~D.}\ \bibnamefont {Reid}},\ and\ \bibinfo {author}
  {\bibfnamefont {P.~K.}\ \bibnamefont {Lam}},\ }\bibfield  {title} {\bibinfo
  {title} {Nonlinear entanglement and its application to generating cat
  states},\ }\href {https://doi.org/10.1103/PhysRevLett.114.100403} {\bibfield
  {journal} {\bibinfo  {journal} {Phys. Rev. Lett.}\ }\textbf {\bibinfo
  {volume} {114}},\ \bibinfo {pages} {100403} (\bibinfo {year}
  {2015})}\BibitemShut {NoStop}%
\bibitem [{\citenamefont {Zhang}\ \emph {et~al.}(2021)\citenamefont {Zhang},
  \citenamefont {Barral}, \citenamefont {Cai}, \citenamefont {Zhang},
  \citenamefont {Xiao},\ and\ \citenamefont {Bencheikh}}]{zhangda2021prl}%
  \BibitemOpen
  \bibfield  {author} {\bibinfo {author} {\bibfnamefont {D.}~\bibnamefont
  {Zhang}}, \bibinfo {author} {\bibfnamefont {D.}~\bibnamefont {Barral}},
  \bibinfo {author} {\bibfnamefont {Y.}~\bibnamefont {Cai}}, \bibinfo {author}
  {\bibfnamefont {Y.}~\bibnamefont {Zhang}}, \bibinfo {author} {\bibfnamefont
  {M.}~\bibnamefont {Xiao}},\ and\ \bibinfo {author} {\bibfnamefont
  {K.}~\bibnamefont {Bencheikh}},\ }\bibfield  {title} {\bibinfo {title}
  {Hierarchy of nonlinear entanglement dynamics for continuous variables},\
  }\href {https://doi.org/10.1103/PhysRevLett.127.150502} {\bibfield  {journal}
  {\bibinfo  {journal} {Phys. Rev. Lett.}\ }\textbf {\bibinfo {volume} {127}},\
  \bibinfo {pages} {150502} (\bibinfo {year} {2021})}\BibitemShut {NoStop}%
\bibitem [{\citenamefont {Lopetegui}\ \emph {et~al.}(2022)\citenamefont
  {Lopetegui}, \citenamefont {Gessner}, \citenamefont {Fadel}, \citenamefont
  {Treps},\ and\ \citenamefont {Walschaers}}]{PRXQuantum.3.030347}%
  \BibitemOpen
  \bibfield  {author} {\bibinfo {author} {\bibfnamefont {C.~E.}\ \bibnamefont
  {Lopetegui}}, \bibinfo {author} {\bibfnamefont {M.}~\bibnamefont {Gessner}},
  \bibinfo {author} {\bibfnamefont {M.}~\bibnamefont {Fadel}}, \bibinfo
  {author} {\bibfnamefont {N.}~\bibnamefont {Treps}},\ and\ \bibinfo {author}
  {\bibfnamefont {M.}~\bibnamefont {Walschaers}},\ }\bibfield  {title}
  {\bibinfo {title} {Homodyne detection of non-{G}aussian quantum steering},\
  }\href {https://doi.org/10.1103/PRXQuantum.3.030347} {\bibfield  {journal}
  {\bibinfo  {journal} {PRX Quantum}\ }\textbf {\bibinfo {volume} {3}},\
  \bibinfo {pages} {030347} (\bibinfo {year} {2022})}\BibitemShut {NoStop}%
\bibitem [{\citenamefont {Yadin}\ \emph {et~al.}(2021)\citenamefont {Yadin},
  \citenamefont {Fadel},\ and\ \citenamefont {Gessner}}]{NC2021}%
  \BibitemOpen
  \bibfield  {author} {\bibinfo {author} {\bibfnamefont {B.}~\bibnamefont
  {Yadin}}, \bibinfo {author} {\bibfnamefont {M.}~\bibnamefont {Fadel}},\ and\
  \bibinfo {author} {\bibfnamefont {M.}~\bibnamefont {Gessner}},\ }\bibfield
  {title} {\bibinfo {title} {Metrological complementarity reveals the
  {E}instein-{P}odolsky-{R}osen paradox},\ }\href
  {https://doi.org/10.1038/s41467-021-22353-3} {\bibfield  {journal} {\bibinfo
  {journal} {Nat. Commun.}\ }\textbf {\bibinfo {volume} {12}},\ \bibinfo
  {pages} {2410} (\bibinfo {year} {2021})}\BibitemShut {NoStop}%
\bibitem [{\citenamefont {Tian}\ \emph {et~al.}(2022)\citenamefont {Tian},
  \citenamefont {Xiang}, \citenamefont {Sun}, \citenamefont {Fadel},\ and\
  \citenamefont {He}}]{tian2022praaplied}%
  \BibitemOpen
  \bibfield  {author} {\bibinfo {author} {\bibfnamefont {M.}~\bibnamefont
  {Tian}}, \bibinfo {author} {\bibfnamefont {Y.}~\bibnamefont {Xiang}},
  \bibinfo {author} {\bibfnamefont {F.-X.}\ \bibnamefont {Sun}}, \bibinfo
  {author} {\bibfnamefont {M.}~\bibnamefont {Fadel}},\ and\ \bibinfo {author}
  {\bibfnamefont {Q.}~\bibnamefont {He}},\ }\bibfield  {title} {\bibinfo
  {title} {Characterizing multipartite non-{G}aussian entanglement for
  three-mode spontaneous parametric down-conversion process},\ }\href
  {https://doi.org/10.1103/PhysRevApplied.18.024065} {\bibfield  {journal}
  {\bibinfo  {journal} {Phys. Rev. Appl.}\ }\textbf {\bibinfo {volume} {18}},\
  \bibinfo {pages} {024065} (\bibinfo {year} {2022})}\BibitemShut {NoStop}%
\bibitem [{\citenamefont {Parigi}\ \emph {et~al.}(2007)\citenamefont {Parigi},
  \citenamefont {Zavatta}, \citenamefont {Kim},\ and\ \citenamefont
  {Bellini}}]{subtraction2007science}%
  \BibitemOpen
  \bibfield  {author} {\bibinfo {author} {\bibfnamefont {V.}~\bibnamefont
  {Parigi}}, \bibinfo {author} {\bibfnamefont {A.}~\bibnamefont {Zavatta}},
  \bibinfo {author} {\bibfnamefont {M.}~\bibnamefont {Kim}},\ and\ \bibinfo
  {author} {\bibfnamefont {M.}~\bibnamefont {Bellini}},\ }\bibfield  {title}
  {\bibinfo {title} {Probing quantum commutation rules by addition and
  subtraction of single photons to/from a light field},\ }\href
  {https://doi.org/10.1126/science.1146204} {\bibfield  {journal} {\bibinfo
  {journal} {Science}\ }\textbf {\bibinfo {volume} {317}},\ \bibinfo {pages}
  {1890} (\bibinfo {year} {2007})}\BibitemShut {NoStop}%
\bibitem [{\citenamefont {Ra}\ \emph {et~al.}(2020)\citenamefont {Ra},
  \citenamefont {Dufour}, \citenamefont {Walschaers}, \citenamefont {Jacquard},
  \citenamefont {Michel}, \citenamefont {Fabre},\ and\ \citenamefont
  {Treps}}]{nicolasnp2020}%
  \BibitemOpen
  \bibfield  {author} {\bibinfo {author} {\bibfnamefont {Y.-S.}\ \bibnamefont
  {Ra}}, \bibinfo {author} {\bibfnamefont {A.}~\bibnamefont {Dufour}}, \bibinfo
  {author} {\bibfnamefont {M.}~\bibnamefont {Walschaers}}, \bibinfo {author}
  {\bibfnamefont {C.}~\bibnamefont {Jacquard}}, \bibinfo {author}
  {\bibfnamefont {T.}~\bibnamefont {Michel}}, \bibinfo {author} {\bibfnamefont
  {C.}~\bibnamefont {Fabre}},\ and\ \bibinfo {author} {\bibfnamefont
  {N.}~\bibnamefont {Treps}},\ }\bibfield  {title} {\bibinfo {title}
  {Non-{G}aussian quantum states of a multimode light field},\ }\href
  {https://doi.org/10.1038/s41567-019-0726-y} {\bibfield  {journal} {\bibinfo
  {journal} {Nat. Phys.}\ }\textbf {\bibinfo {volume} {16}},\ \bibinfo {pages}
  {144} (\bibinfo {year} {2020})}\BibitemShut {NoStop}%
\bibitem [{\citenamefont {Mirrahimi}\ \emph {et~al.}(2014)\citenamefont
  {Mirrahimi}, \citenamefont {Leghtas}, \citenamefont {Albert}, \citenamefont
  {Touzard}, \citenamefont {Schoelkopf}, \citenamefont {Jiang},\ and\
  \citenamefont {Devoret}}]{error-2014}%
  \BibitemOpen
  \bibfield  {author} {\bibinfo {author} {\bibfnamefont {M.}~\bibnamefont
  {Mirrahimi}}, \bibinfo {author} {\bibfnamefont {Z.}~\bibnamefont {Leghtas}},
  \bibinfo {author} {\bibfnamefont {V.~V.}\ \bibnamefont {Albert}}, \bibinfo
  {author} {\bibfnamefont {S.}~\bibnamefont {Touzard}}, \bibinfo {author}
  {\bibfnamefont {R.~J.}\ \bibnamefont {Schoelkopf}}, \bibinfo {author}
  {\bibfnamefont {L.}~\bibnamefont {Jiang}},\ and\ \bibinfo {author}
  {\bibfnamefont {M.~H.}\ \bibnamefont {Devoret}},\ }\bibfield  {title}
  {\bibinfo {title} {Dynamically protected cat-qubits: a new paradigm for
  universal quantum computation},\ }\href
  {https://doi.org/10.1088/1367-2630/16/4/045014} {\bibfield  {journal}
  {\bibinfo  {journal} {New J. Phys.}\ }\textbf {\bibinfo {volume} {16}},\
  \bibinfo {pages} {045014} (\bibinfo {year} {2014})}\BibitemShut {NoStop}%
\bibitem [{\citenamefont {Li}\ \emph {et~al.}(2017)\citenamefont {Li},
  \citenamefont {Zou}, \citenamefont {Albert}, \citenamefont {Muralidharan},
  \citenamefont {Girvin},\ and\ \citenamefont
  {Jiang}}]{error-PhysRevLett.119.030502}%
  \BibitemOpen
  \bibfield  {author} {\bibinfo {author} {\bibfnamefont {L.}~\bibnamefont
  {Li}}, \bibinfo {author} {\bibfnamefont {C.-L.}\ \bibnamefont {Zou}},
  \bibinfo {author} {\bibfnamefont {V.~V.}\ \bibnamefont {Albert}}, \bibinfo
  {author} {\bibfnamefont {S.}~\bibnamefont {Muralidharan}}, \bibinfo {author}
  {\bibfnamefont {S.~M.}\ \bibnamefont {Girvin}},\ and\ \bibinfo {author}
  {\bibfnamefont {L.}~\bibnamefont {Jiang}},\ }\bibfield  {title} {\bibinfo
  {title} {Cat codes with optimal decoherence suppression for a lossy bosonic
  channel},\ }\href {https://doi.org/10.1103/PhysRevLett.119.030502} {\bibfield
   {journal} {\bibinfo  {journal} {Phys. Rev. Lett.}\ }\textbf {\bibinfo
  {volume} {119}},\ \bibinfo {pages} {030502} (\bibinfo {year}
  {2017})}\BibitemShut {NoStop}%
\bibitem [{\citenamefont {Cai}\ \emph {et~al.}(2021)\citenamefont {Cai},
  \citenamefont {Ma}, \citenamefont {Wang}, \citenamefont {Zou},\ and\
  \citenamefont {Sun}}]{errorcorrection2021}%
  \BibitemOpen
  \bibfield  {author} {\bibinfo {author} {\bibfnamefont {W.}~\bibnamefont
  {Cai}}, \bibinfo {author} {\bibfnamefont {Y.}~\bibnamefont {Ma}}, \bibinfo
  {author} {\bibfnamefont {W.}~\bibnamefont {Wang}}, \bibinfo {author}
  {\bibfnamefont {C.-L.}\ \bibnamefont {Zou}},\ and\ \bibinfo {author}
  {\bibfnamefont {L.}~\bibnamefont {Sun}},\ }\bibfield  {title} {\bibinfo
  {title} {Bosonic quantum error correction codes in superconducting quantum
  circuits},\ }\href
  {https://doi.org/https://doi.org/10.1016/j.fmre.2020.12.006} {\bibfield
  {journal} {\bibinfo  {journal} {Fundam. Res.}\ }\textbf {\bibinfo {volume}
  {1}},\ \bibinfo {pages} {50} (\bibinfo {year} {2021})}\BibitemShut {NoStop}%
\bibitem [{\citenamefont {Werner}\ and\ \citenamefont
  {Wolf}(2001)}]{boundentanglement}%
  \BibitemOpen
  \bibfield  {author} {\bibinfo {author} {\bibfnamefont {R.~F.}\ \bibnamefont
  {Werner}}\ and\ \bibinfo {author} {\bibfnamefont {M.~M.}\ \bibnamefont
  {Wolf}},\ }\bibfield  {title} {\bibinfo {title} {Bound entangled {G}aussian
  states},\ }\href {https://doi.org/10.1103/PhysRevLett.86.3658} {\bibfield
  {journal} {\bibinfo  {journal} {Phys. Rev. Lett.}\ }\textbf {\bibinfo
  {volume} {86}},\ \bibinfo {pages} {3658} (\bibinfo {year}
  {2001})}\BibitemShut {NoStop}%
\bibitem [{\citenamefont {Hillery}(1987)}]{hillerypra1987}%
  \BibitemOpen
  \bibfield  {author} {\bibinfo {author} {\bibfnamefont {M.}~\bibnamefont
  {Hillery}},\ }\bibfield  {title} {\bibinfo {title} {Amplitude-squared
  squeezing of the electromagnetic field},\ }\href
  {https://doi.org/10.1103/PhysRevA.36.3796} {\bibfield  {journal} {\bibinfo
  {journal} {Phys. Rev. A}\ }\textbf {\bibinfo {volume} {36}},\ \bibinfo
  {pages} {3796} (\bibinfo {year} {1987})}\BibitemShut {NoStop}%
\bibitem [{\citenamefont {Tripathi}\ \emph {et~al.}(2020)\citenamefont
  {Tripathi}, \citenamefont {Radhakrishnan},\ and\ \citenamefont
  {Byrnes}}]{njp-entanglement}%
  \BibitemOpen
  \bibfield  {author} {\bibinfo {author} {\bibfnamefont {V.}~\bibnamefont
  {Tripathi}}, \bibinfo {author} {\bibfnamefont {C.}~\bibnamefont
  {Radhakrishnan}},\ and\ \bibinfo {author} {\bibfnamefont {T.}~\bibnamefont
  {Byrnes}},\ }\bibfield  {title} {\bibinfo {title} {Covariance matrix
  entanglement criterion for an arbitrary set of operators},\ }\href
  {https://doi.org/10.1088/1367-2630/ab9ce7} {\bibfield  {journal} {\bibinfo
  {journal} {New J. Phys.}\ }\textbf {\bibinfo {volume} {22}},\ \bibinfo
  {pages} {073055} (\bibinfo {year} {2020})}\BibitemShut {NoStop}%
\bibitem [{\citenamefont {Ivan}\ \emph {et~al.}(2012)\citenamefont {Ivan},
  \citenamefont {Sabapathy}, \citenamefont {Mukunda},\ and\ \citenamefont
  {Simon}}]{cm-high}%
  \BibitemOpen
  \bibfield  {author} {\bibinfo {author} {\bibfnamefont {J.~S.}\ \bibnamefont
  {Ivan}}, \bibinfo {author} {\bibfnamefont {K.~K.}\ \bibnamefont {Sabapathy}},
  \bibinfo {author} {\bibfnamefont {N.}~\bibnamefont {Mukunda}},\ and\ \bibinfo
  {author} {\bibfnamefont {R.}~\bibnamefont {Simon}},\ }\bibfield  {title}
  {\bibinfo {title} {Invariant theoretic approach to uncertainty relations for
  quantum systems},\ }\href {https://doi.org/10.48550/arXiv.1205.5132}
  {\bibfield  {journal} {\bibinfo  {journal} {arXiv:1205.5132}\ } (\bibinfo
  {year} {2012})}\BibitemShut {NoStop}%
\bibitem [{\citenamefont {Lee}\ \emph {et~al.}(2014)\citenamefont {Lee},
  \citenamefont {Ryu}, \citenamefont {Bang},\ and\ \citenamefont {Nha}}]{Nha}%
  \BibitemOpen
  \bibfield  {author} {\bibinfo {author} {\bibfnamefont {C.-W.}\ \bibnamefont
  {Lee}}, \bibinfo {author} {\bibfnamefont {J.}~\bibnamefont {Ryu}}, \bibinfo
  {author} {\bibfnamefont {J.}~\bibnamefont {Bang}},\ and\ \bibinfo {author}
  {\bibfnamefont {H.}~\bibnamefont {Nha}},\ }\bibfield  {title} {\bibinfo
  {title} {Inseparability criterion using higher-order
  {S}chr\"{o}dinger-{R}obertson uncertainty relation},\ }\href
  {https://doi.org/10.1364/JOSAB.31.000656} {\bibfield  {journal} {\bibinfo
  {journal} {J. Opt. Soc. Am. B}\ }\textbf {\bibinfo {volume} {31}},\ \bibinfo
  {pages} {656} (\bibinfo {year} {2014})}\BibitemShut {NoStop}%
\bibitem [{\citenamefont {Ji}\ \emph {et~al.}(2015)\citenamefont {Ji},
  \citenamefont {Lee}, \citenamefont {Park},\ and\ \citenamefont
  {Nha}}]{eprcm-pra2014}%
  \BibitemOpen
  \bibfield  {author} {\bibinfo {author} {\bibfnamefont {S.-W.}\ \bibnamefont
  {Ji}}, \bibinfo {author} {\bibfnamefont {J.}~\bibnamefont {Lee}}, \bibinfo
  {author} {\bibfnamefont {J.}~\bibnamefont {Park}},\ and\ \bibinfo {author}
  {\bibfnamefont {H.}~\bibnamefont {Nha}},\ }\bibfield  {title} {\bibinfo
  {title} {Steering criteria via covariance matrices of local observables in
  arbitrary-dimensional quantum systems},\ }\href
  {https://doi.org/10.1103/PhysRevA.92.062130} {\bibfield  {journal} {\bibinfo
  {journal} {Phys. Rev. A}\ }\textbf {\bibinfo {volume} {92}},\ \bibinfo
  {pages} {062130} (\bibinfo {year} {2015})}\BibitemShut {NoStop}%
\bibitem [{\citenamefont {Cavalcanti}\ \emph {et~al.}(2011)\citenamefont
  {Cavalcanti}, \citenamefont {He}, \citenamefont {Reid},\ and\ \citenamefont
  {Wiseman}}]{hzhe2011}%
  \BibitemOpen
  \bibfield  {author} {\bibinfo {author} {\bibfnamefont {E.~G.}\ \bibnamefont
  {Cavalcanti}}, \bibinfo {author} {\bibfnamefont {Q.~Y.}\ \bibnamefont {He}},
  \bibinfo {author} {\bibfnamefont {M.~D.}\ \bibnamefont {Reid}},\ and\
  \bibinfo {author} {\bibfnamefont {H.~M.}\ \bibnamefont {Wiseman}},\
  }\bibfield  {title} {\bibinfo {title} {Unified criteria for multipartite
  quantum nonlocality},\ }\href {https://doi.org/10.1103/PhysRevA.84.032115}
  {\bibfield  {journal} {\bibinfo  {journal} {Phys. Rev. A}\ }\textbf {\bibinfo
  {volume} {84}},\ \bibinfo {pages} {032115} (\bibinfo {year}
  {2011})}\BibitemShut {NoStop}%
\bibitem [{\citenamefont {He}\ \emph {et~al.}(2012)\citenamefont {He},
  \citenamefont {Drummond}, \citenamefont {Olsen},\ and\ \citenamefont
  {Reid}}]{hz-he2012}%
  \BibitemOpen
  \bibfield  {author} {\bibinfo {author} {\bibfnamefont {Q.~Y.}\ \bibnamefont
  {He}}, \bibinfo {author} {\bibfnamefont {P.~D.}\ \bibnamefont {Drummond}},
  \bibinfo {author} {\bibfnamefont {M.~K.}\ \bibnamefont {Olsen}},\ and\
  \bibinfo {author} {\bibfnamefont {M.~D.}\ \bibnamefont {Reid}},\ }\bibfield
  {title} {\bibinfo {title} {Einstein-{P}odolsky-{R}osen entanglement and
  steering in two-well {B}ose-{E}instein-condensate ground states},\ }\href
  {https://doi.org/10.1103/PhysRevA.86.023626} {\bibfield  {journal} {\bibinfo
  {journal} {Phys. Rev. A}\ }\textbf {\bibinfo {volume} {86}},\ \bibinfo
  {pages} {023626} (\bibinfo {year} {2012})}\BibitemShut {NoStop}%
\bibitem [{\citenamefont {Reid}(1989)}]{reid1989demonstration}%
  \BibitemOpen
  \bibfield  {author} {\bibinfo {author} {\bibfnamefont {M.}~\bibnamefont
  {Reid}},\ }\bibfield  {title} {\bibinfo {title} {Demonstration of the
  {E}instein-{P}odolsky-{R}osen paradox using nondegenerate parametric
  amplification},\ }\href {https://doi.org/10.1103/PhysRevA.40.913} {\bibfield
  {journal} {\bibinfo  {journal} {Phys. Rev. A}\ }\textbf {\bibinfo {volume}
  {40}},\ \bibinfo {pages} {913} (\bibinfo {year} {1989})}\BibitemShut
  {NoStop}%
\bibitem [{\citenamefont {Tan}\ \emph {et~al.}(2015)\citenamefont {Tan},
  \citenamefont {Zhang},\ and\ \citenamefont {Li}}]{equiv-tan}%
  \BibitemOpen
  \bibfield  {author} {\bibinfo {author} {\bibfnamefont {H.}~\bibnamefont
  {Tan}}, \bibinfo {author} {\bibfnamefont {X.}~\bibnamefont {Zhang}},\ and\
  \bibinfo {author} {\bibfnamefont {G.}~\bibnamefont {Li}},\ }\bibfield
  {title} {\bibinfo {title} {Steady-state one-way {E}instein-{P}odolsky-{R}osen
  steering in optomechanical interfaces},\ }\href
  {https://doi.org/10.1103/PhysRevA.91.032121} {\bibfield  {journal} {\bibinfo
  {journal} {Phys. Rev. A}\ }\textbf {\bibinfo {volume} {91}},\ \bibinfo
  {pages} {032121} (\bibinfo {year} {2015})}\BibitemShut {NoStop}%
\bibitem [{\citenamefont {Zheng}\ \emph {et~al.}(2019)\citenamefont {Zheng},
  \citenamefont {Sun}, \citenamefont {Lai}, \citenamefont {Gong},\ and\
  \citenamefont {He}}]{equiv-zss}%
  \BibitemOpen
  \bibfield  {author} {\bibinfo {author} {\bibfnamefont {S.}~\bibnamefont
  {Zheng}}, \bibinfo {author} {\bibfnamefont {F.}~\bibnamefont {Sun}}, \bibinfo
  {author} {\bibfnamefont {Y.}~\bibnamefont {Lai}}, \bibinfo {author}
  {\bibfnamefont {Q.}~\bibnamefont {Gong}},\ and\ \bibinfo {author}
  {\bibfnamefont {Q.}~\bibnamefont {He}},\ }\bibfield  {title} {\bibinfo
  {title} {Manipulation and enhancement of asymmetric steering via interference
  effects induced by closed-loop coupling},\ }\href
  {https://doi.org/10.1103/PhysRevA.99.022335} {\bibfield  {journal} {\bibinfo
  {journal} {Phys. Rev. A}\ }\textbf {\bibinfo {volume} {99}},\ \bibinfo
  {pages} {022335} (\bibinfo {year} {2019})}\BibitemShut {NoStop}%
\bibitem [{\citenamefont {Jing}\ \emph {et~al.}(2019)\citenamefont {Jing},
  \citenamefont {Fadel}, \citenamefont {Ivannikov},\ and\ \citenamefont
  {Byrnes}}]{eqiv-spin}%
  \BibitemOpen
  \bibfield  {author} {\bibinfo {author} {\bibfnamefont {Y.}~\bibnamefont
  {Jing}}, \bibinfo {author} {\bibfnamefont {M.}~\bibnamefont {Fadel}},
  \bibinfo {author} {\bibfnamefont {V.}~\bibnamefont {Ivannikov}},\ and\
  \bibinfo {author} {\bibfnamefont {T.}~\bibnamefont {Byrnes}},\ }\bibfield
  {title} {\bibinfo {title} {Split spin-squeezed {B}ose–{E}instein
  condensates},\ }\href {https://doi.org/10.1088/1367-2630/ab3fcf} {\bibfield
  {journal} {\bibinfo  {journal} {New J. Phys.}\ }\textbf {\bibinfo {volume}
  {21}},\ \bibinfo {pages} {093038} (\bibinfo {year} {2019})}\BibitemShut
  {NoStop}%
\bibitem [{\citenamefont {Strang}(2016)}]{linearalgebra}%
  \BibitemOpen
  \bibfield  {author} {\bibinfo {author} {\bibfnamefont {G.}~\bibnamefont
  {Strang}},\ }\bibinfo {title} {Introduction to linear algebra}\ (\bibinfo
  {publisher} {Wellesley-Cambridge Press},\ \bibinfo {year} {2016})\
  Chap.~\bibinfo {chapter} {6}, pp.\ \bibinfo {pages} {350--360},\ \bibinfo
  {edition} {5th}\ ed.\BibitemShut {Stop}%
\bibitem [{\citenamefont {Furusawa}\ \emph {et~al.}(1998)\citenamefont
  {Furusawa}, \citenamefont {Sørensen}, \citenamefont {Braunstein},
  \citenamefont {Fuchs}, \citenamefont {Kimble},\ and\ \citenamefont
  {Polzik}}]{eprstate}%
  \BibitemOpen
  \bibfield  {author} {\bibinfo {author} {\bibfnamefont {A.}~\bibnamefont
  {Furusawa}}, \bibinfo {author} {\bibfnamefont {J.~L.}\ \bibnamefont
  {Sørensen}}, \bibinfo {author} {\bibfnamefont {S.~L.}\ \bibnamefont
  {Braunstein}}, \bibinfo {author} {\bibfnamefont {C.~A.}\ \bibnamefont
  {Fuchs}}, \bibinfo {author} {\bibfnamefont {H.~J.}\ \bibnamefont {Kimble}},\
  and\ \bibinfo {author} {\bibfnamefont {E.~S.}\ \bibnamefont {Polzik}},\
  }\bibfield  {title} {\bibinfo {title} {Unconditional quantum teleportation},\
  }\href {https://doi.org/10.1126/science.282.5389.706} {\bibfield  {journal}
  {\bibinfo  {journal} {Science}\ }\textbf {\bibinfo {volume} {282}},\ \bibinfo
  {pages} {706} (\bibinfo {year} {1998})}\BibitemShut {NoStop}%
\bibitem [{\citenamefont {Sun}\ \emph {et~al.}(2021)\citenamefont {Sun},
  \citenamefont {Zheng}, \citenamefont {Xiao}, \citenamefont {Gong},
  \citenamefont {He},\ and\ \citenamefont {Xia}}]{sunfengxiao2021prl}%
  \BibitemOpen
  \bibfield  {author} {\bibinfo {author} {\bibfnamefont {F.-X.}\ \bibnamefont
  {Sun}}, \bibinfo {author} {\bibfnamefont {S.-S.}\ \bibnamefont {Zheng}},
  \bibinfo {author} {\bibfnamefont {Y.}~\bibnamefont {Xiao}}, \bibinfo {author}
  {\bibfnamefont {Q.}~\bibnamefont {Gong}}, \bibinfo {author} {\bibfnamefont
  {Q.}~\bibnamefont {He}},\ and\ \bibinfo {author} {\bibfnamefont
  {K.}~\bibnamefont {Xia}},\ }\bibfield  {title} {\bibinfo {title} {Remote
  generation of magnon {S}chr\"odinger cat state via magnon-photon
  entanglement},\ }\href {https://doi.org/10.1103/PhysRevLett.127.087203}
  {\bibfield  {journal} {\bibinfo  {journal} {Phys. Rev. Lett.}\ }\textbf
  {\bibinfo {volume} {127}},\ \bibinfo {pages} {087203} (\bibinfo {year}
  {2021})}\BibitemShut {NoStop}%
\bibitem [{\citenamefont {Ourjoumtsev}\ \emph {et~al.}(2006)\citenamefont
  {Ourjoumtsev}, \citenamefont {Tualle-Brouri}, \citenamefont {Laurat},\ and\
  \citenamefont {Grangier}}]{cat2006}%
  \BibitemOpen
  \bibfield  {author} {\bibinfo {author} {\bibfnamefont {A.}~\bibnamefont
  {Ourjoumtsev}}, \bibinfo {author} {\bibfnamefont {R.}~\bibnamefont
  {Tualle-Brouri}}, \bibinfo {author} {\bibfnamefont {J.}~\bibnamefont
  {Laurat}},\ and\ \bibinfo {author} {\bibfnamefont {P.}~\bibnamefont
  {Grangier}},\ }\bibfield  {title} {\bibinfo {title} {Generating optical
  {S}chrödinger kittens for quantum information processing},\ }\href
  {https://doi.org/10.1126/science.1122858} {\bibfield  {journal} {\bibinfo
  {journal} {Science}\ }\textbf {\bibinfo {volume} {312}},\ \bibinfo {pages}
  {83} (\bibinfo {year} {2006})}\BibitemShut {NoStop}%
\bibitem [{\citenamefont {Ourjoumtsev}\ \emph
  {et~al.}(2007{\natexlab{a}})\citenamefont {Ourjoumtsev}, \citenamefont
  {Jeong}, \citenamefont {Tualle-Brouri},\ and\ \citenamefont
  {Grangier}}]{cat2007}%
  \BibitemOpen
  \bibfield  {author} {\bibinfo {author} {\bibfnamefont {A.}~\bibnamefont
  {Ourjoumtsev}}, \bibinfo {author} {\bibfnamefont {H.}~\bibnamefont {Jeong}},
  \bibinfo {author} {\bibfnamefont {R.}~\bibnamefont {Tualle-Brouri}},\ and\
  \bibinfo {author} {\bibfnamefont {P.}~\bibnamefont {Grangier}},\ }\bibfield
  {title} {\bibinfo {title} {Generation of optical `{S}chrödinger cats' from
  photon number states},\ }\href {https://doi.org/10.1038/nature06054}
  {\bibfield  {journal} {\bibinfo  {journal} {Nature}\ }\textbf {\bibinfo
  {volume} {448}},\ \bibinfo {pages} {784} (\bibinfo {year}
  {2007}{\natexlab{a}})}\BibitemShut {NoStop}%
\bibitem [{\citenamefont {Lvovsky}\ \emph {et~al.}(2013)\citenamefont
  {Lvovsky}, \citenamefont {Ghobadi}, \citenamefont {Chandra}, \citenamefont
  {Prasad},\ and\ \citenamefont {Simon}}]{cat2013}%
  \BibitemOpen
  \bibfield  {author} {\bibinfo {author} {\bibfnamefont {A.~I.}\ \bibnamefont
  {Lvovsky}}, \bibinfo {author} {\bibfnamefont {R.}~\bibnamefont {Ghobadi}},
  \bibinfo {author} {\bibfnamefont {A.}~\bibnamefont {Chandra}}, \bibinfo
  {author} {\bibfnamefont {A.~S.}\ \bibnamefont {Prasad}},\ and\ \bibinfo
  {author} {\bibfnamefont {C.}~\bibnamefont {Simon}},\ }\bibfield  {title}
  {\bibinfo {title} {Observation of micro–macro entanglement of light},\
  }\href {https://doi.org/10.1038/nphys2682} {\bibfield  {journal} {\bibinfo
  {journal} {Nat. Phys.}\ }\textbf {\bibinfo {volume} {9}},\ \bibinfo {pages}
  {541} (\bibinfo {year} {2013})}\BibitemShut {NoStop}%
\bibitem [{\citenamefont {Ourjoumtsev}\ \emph {et~al.}(2009)\citenamefont
  {Ourjoumtsev}, \citenamefont {Ferreyrol}, \citenamefont {Tualle-Brouri},\
  and\ \citenamefont {Grangier}}]{subtraction-np2009}%
  \BibitemOpen
  \bibfield  {author} {\bibinfo {author} {\bibfnamefont {A.}~\bibnamefont
  {Ourjoumtsev}}, \bibinfo {author} {\bibfnamefont {F.}~\bibnamefont
  {Ferreyrol}}, \bibinfo {author} {\bibfnamefont {R.}~\bibnamefont
  {Tualle-Brouri}},\ and\ \bibinfo {author} {\bibfnamefont {P.}~\bibnamefont
  {Grangier}},\ }\bibfield  {title} {\bibinfo {title} {Preparation of non-local
  superpositions of quasi-classical light states},\ }\href
  {https://doi.org/10.1038/nphys1199} {\bibfield  {journal} {\bibinfo
  {journal} {Nat. Phys.}\ }\textbf {\bibinfo {volume} {5}},\ \bibinfo {pages}
  {189} (\bibinfo {year} {2009})}\BibitemShut {NoStop}%
\bibitem [{\citenamefont {Ourjoumtsev}\ \emph
  {et~al.}(2007{\natexlab{b}})\citenamefont {Ourjoumtsev}, \citenamefont
  {Dantan}, \citenamefont {Tualle-Brouri},\ and\ \citenamefont
  {Grangier}}]{subtraction-PhysRevLett.98.030502}%
  \BibitemOpen
  \bibfield  {author} {\bibinfo {author} {\bibfnamefont {A.}~\bibnamefont
  {Ourjoumtsev}}, \bibinfo {author} {\bibfnamefont {A.}~\bibnamefont {Dantan}},
  \bibinfo {author} {\bibfnamefont {R.}~\bibnamefont {Tualle-Brouri}},\ and\
  \bibinfo {author} {\bibfnamefont {P.}~\bibnamefont {Grangier}},\ }\bibfield
  {title} {\bibinfo {title} {Increasing entanglement between {G}aussian states
  by coherent photon subtraction},\ }\href
  {https://doi.org/10.1103/PhysRevLett.98.030502} {\bibfield  {journal}
  {\bibinfo  {journal} {Phys. Rev. Lett.}\ }\textbf {\bibinfo {volume} {98}},\
  \bibinfo {pages} {030502} (\bibinfo {year} {2007}{\natexlab{b}})}\BibitemShut
  {NoStop}%
\bibitem [{\citenamefont {Xiang}\ \emph {et~al.}(2017)\citenamefont {Xiang},
  \citenamefont {Xu}, \citenamefont {Mi\ifmmode~\check{s}\else \v{s}\fi{}ta},
  \citenamefont {Tufarelli}, \citenamefont {He},\ and\ \citenamefont
  {Adesso}}]{xiangyu2017}%
  \BibitemOpen
  \bibfield  {author} {\bibinfo {author} {\bibfnamefont {Y.}~\bibnamefont
  {Xiang}}, \bibinfo {author} {\bibfnamefont {B.}~\bibnamefont {Xu}}, \bibinfo
  {author} {\bibfnamefont {L.}~\bibnamefont {Mi\ifmmode~\check{s}\else
  \v{s}\fi{}ta}}, \bibinfo {author} {\bibfnamefont {T.}~\bibnamefont
  {Tufarelli}}, \bibinfo {author} {\bibfnamefont {Q.}~\bibnamefont {He}},\ and\
  \bibinfo {author} {\bibfnamefont {G.}~\bibnamefont {Adesso}},\ }\bibfield
  {title} {\bibinfo {title} {Investigating {E}instein-{P}odolsky-{R}osen
  steering of continuous-variable bipartite states by non-{G}aussian pseudospin
  measurements},\ }\href {https://doi.org/10.1103/PhysRevA.96.042326}
  {\bibfield  {journal} {\bibinfo  {journal} {Phys. Rev. A}\ }\textbf {\bibinfo
  {volume} {96}},\ \bibinfo {pages} {042326} (\bibinfo {year}
  {2017})}\BibitemShut {NoStop}%
\bibitem [{\citenamefont {Flühmann}\ \emph {et~al.}(2019)\citenamefont
  {Flühmann}, \citenamefont {Nguyen}, \citenamefont {Marinelli}, \citenamefont
  {Negnevitsky}, \citenamefont {Mehta},\ and\ \citenamefont {Home}}]{ion-ref}%
  \BibitemOpen
  \bibfield  {author} {\bibinfo {author} {\bibfnamefont {C.}~\bibnamefont
  {Flühmann}}, \bibinfo {author} {\bibfnamefont {T.~L.}\ \bibnamefont
  {Nguyen}}, \bibinfo {author} {\bibfnamefont {M.}~\bibnamefont {Marinelli}},
  \bibinfo {author} {\bibfnamefont {V.}~\bibnamefont {Negnevitsky}}, \bibinfo
  {author} {\bibfnamefont {K.}~\bibnamefont {Mehta}},\ and\ \bibinfo {author}
  {\bibfnamefont {J.~P.}\ \bibnamefont {Home}},\ }\bibfield  {title} {\bibinfo
  {title} {Encoding a qubit in a trapped-ion mechanical oscillator},\ }\href
  {https://doi.org/10.1038/s41586-019-0960-6} {\bibfield  {journal} {\bibinfo
  {journal} {Nature}\ }\textbf {\bibinfo {volume} {566}},\ \bibinfo {pages}
  {513} (\bibinfo {year} {2019})}\BibitemShut {NoStop}%
\bibitem [{\citenamefont {Jost}\ \emph {et~al.}(2009)\citenamefont {Jost},
  \citenamefont {Home}, \citenamefont {Amini}, \citenamefont {Hanneke},
  \citenamefont {Ozeri}, \citenamefont {Langer}, \citenamefont {Bollinger},
  \citenamefont {Leibfried},\ and\ \citenamefont {Wineland}}]{opomech2009}%
  \BibitemOpen
  \bibfield  {author} {\bibinfo {author} {\bibfnamefont {J.~D.}\ \bibnamefont
  {Jost}}, \bibinfo {author} {\bibfnamefont {J.~P.}\ \bibnamefont {Home}},
  \bibinfo {author} {\bibfnamefont {J.~M.}\ \bibnamefont {Amini}}, \bibinfo
  {author} {\bibfnamefont {D.}~\bibnamefont {Hanneke}}, \bibinfo {author}
  {\bibfnamefont {R.}~\bibnamefont {Ozeri}}, \bibinfo {author} {\bibfnamefont
  {C.}~\bibnamefont {Langer}}, \bibinfo {author} {\bibfnamefont {J.~J.}\
  \bibnamefont {Bollinger}}, \bibinfo {author} {\bibfnamefont {D.}~\bibnamefont
  {Leibfried}},\ and\ \bibinfo {author} {\bibfnamefont {D.~J.}\ \bibnamefont
  {Wineland}},\ }\bibfield  {title} {\bibinfo {title} {Entangled mechanical
  oscillators},\ }\href {https://doi.org/10.1038/nature08006} {\bibfield
  {journal} {\bibinfo  {journal} {Nature}\ }\textbf {\bibinfo {volume} {459}},\
  \bibinfo {pages} {683} (\bibinfo {year} {2009})}\BibitemShut {NoStop}%
\bibitem [{\citenamefont {Riedinger}\ \emph {et~al.}(2016)\citenamefont
  {Riedinger}, \citenamefont {Hong}, \citenamefont {Norte}, \citenamefont
  {Slater}, \citenamefont {Shang}, \citenamefont {Krause}, \citenamefont
  {Anant}, \citenamefont {Aspelmeyer},\ and\ \citenamefont
  {Gröblacher}}]{opomech2016}%
  \BibitemOpen
  \bibfield  {author} {\bibinfo {author} {\bibfnamefont {R.}~\bibnamefont
  {Riedinger}}, \bibinfo {author} {\bibfnamefont {S.}~\bibnamefont {Hong}},
  \bibinfo {author} {\bibfnamefont {R.~A.}\ \bibnamefont {Norte}}, \bibinfo
  {author} {\bibfnamefont {J.~A.}\ \bibnamefont {Slater}}, \bibinfo {author}
  {\bibfnamefont {J.}~\bibnamefont {Shang}}, \bibinfo {author} {\bibfnamefont
  {A.~G.}\ \bibnamefont {Krause}}, \bibinfo {author} {\bibfnamefont
  {V.}~\bibnamefont {Anant}}, \bibinfo {author} {\bibfnamefont
  {M.}~\bibnamefont {Aspelmeyer}},\ and\ \bibinfo {author} {\bibfnamefont
  {S.}~\bibnamefont {Gröblacher}},\ }\bibfield  {title} {\bibinfo {title}
  {Non-classical correlations between single photons and phonons from a
  mechanical oscillator},\ }\href {https://doi.org/10.1038/nature16536}
  {\bibfield  {journal} {\bibinfo  {journal} {Nature}\ }\textbf {\bibinfo
  {volume} {530}},\ \bibinfo {pages} {313} (\bibinfo {year}
  {2016})}\BibitemShut {NoStop}%
\end{thebibliography}

%


\end{document}